%% file: main.tex
\documentclass[acmsmall, nonacm]{acmart}

\usepackage{tikz}
\usetikzlibrary{shapes.geometric, arrows, trees, positioning, shapes}
\tikzstyle{process} = [rectangle, minimum width=3cm, minimum height=1cm, text centered, draw=black, fill=gray!15]
\tikzstyle{smallprocess} = [process, minimum width=2cm, minimum height=0.5cm]
\tikzstyle{smallbox} = [rectangle, minimum width=2cm, minimum height=1cm, text centered, draw=black, fill=gray!15]
\tikzstyle{verification} = [rectangle, minimum width=3cm, minimum height=1cm, text centered, draw=black, fill=blue!10]
\tikzstyle{decision} = [diamond, minimum width=5cm, minimum height=1cm, aspect=2, text centered, draw=black, fill=orange!20]
\tikzstyle{roundedbox} = [rectangle, rounded corners, minimum width=3cm, minimum height=1cm, text centered, draw=black]
\tikzstyle{arrow} = [thick,->,>=stealth]
\tikzstyle{dottedarrow} = [thick,->,>=stealth, dotted]

\usepackage{natbib}
\usepackage{tabularx}
\usepackage{paralist}
\usepackage[table]{xcolor}

\usepackage{amsthm}
\theoremstyle{definition}
\newtheorem{definition}{Definition}
\usepackage{framed}

\usepackage{xcolor}

\newcommand{\plabel}[1]{\hypertarget{#1}{#1}}
\newcommand{\pref}[1]{\hyperlink{#1}{[#1]}}

\setcopyright{none}
\renewcommand\footnotetextcopyrightpermission[1]{} %
\begin{document}

\title{A Systematic Mapping Study on the Debugging of Autonomous Driving Systems}

\author{Nathan Shaw}
\affiliation{%
\institution{University of Sheffield}
\department{School of Computer Science}
\city{Sheffield}
\country{United Kingdom}}
\email{Nshaw1@sheffield.ac.uk}
\orcid{0009-0006-6091-2214}

\author{Sanjeetha Pennada}
\affiliation{%
\institution{University of Sheffield}
\department{School of Computer Science}
\city{Sheffield}
\country{United Kingdom}}
\orcid{0000-0003-4650-862X}

\author{Robert M Hierons}
\affiliation{%
\institution{University of Sheffield}
\department{School of Computer Science}
\city{Sheffield}
\country{United Kingdom}}
\orcid{0000-0002-4771-1446}

\author{Donghwan Shin}
\affiliation{%
\institution{University of Sheffield}
\department{School of Computer Science}
\city{Sheffield}
\country{United Kingdom}}
\orcid{0000-0002-0840-6449}

\begin{abstract}
As Autonomous Driving Systems (ADS) progress towards commercial deployment, there is an increasing focus on ensuring their safety and reliability. While considerable research has been conducted on testing methods for detecting faults in ADS, very little attention has been paid to \textit{debugging} in ADS. Debugging is an essential process that follows test failures to localise and repair the faults in the systems to maintain their safety and reliability. This Systematic Mapping Study (SMS) aims to provide a detailed overview of the current landscape of ADS debugging, highlighting existing approaches and identifying gaps in research. The study also proposes directions for future work and standards for problem definition and terminology in the field. Our findings reveal various methods for ADS debugging and highlight the current fragmented yet promising landscape.
\end{abstract}



\maketitle

\input{Sections/Introduction}

\input{Sections/Background}

\input{Sections/Methodology}

\input{Sections/Research_Types}

\input{Sections/Problems}

\input{Sections/Techniques}

\input{Sections/Tools}

\input{Sections/Discussion}

\input{Sections/Conclusion}

\begin{acks}
    This work was supported by the Institute of Information \& Communications Technology Planning \& Evaluation(IITP) grant funded by the Korea government(MSIT) (No. RS-2025-02218761, 50\%) and by the Engineering and Physical Sciences Research Council (EPSRC) [EP/Y014219/1].
\end{acks}

\bibliographystyle{ACM-Reference-Format}
\bibliography{references.bib}

\end{document}

%% file: Sections/Introduction.tex
\section{Introduction}

Autonomous Driving Systems (ADS), the crux of autonomous vehicles, have made vast improvements over the past decade. These improvements reflect growing demand for automated vehicles and their potential to improve traffic safety, reduce emissions and enhanced reliability. These advancements have led some governments, such as the UK government \cite{gov}, to begin preparing regulations for their deployment, as ADS and advanced driving assistance systems (ADAS) are expected to play a major role in the future of transportation.

ADS is a safety-critical application of machine learning-based software systems, with the performance of the system directly impacting the well-being of those who use it. As a result, rigorous testing is essential to ensure the system behaves as expected and to identify any unsafe or incorrect behaviours before deployment. The need to ensure safety has made testing a prominent research topic with many existing surveys detailing the advances in test generation, scenario coverage and other testing methodologies for ADS \cite{testing-review,Tang2023,Koopman2016}. Defined testing protocols ensure that ADS systems are not only functionally correct but adhere to safety, ethical and legal standards required of them \cite{testing-review,Tang2023,Koopman2016}.

Testing, i.e. identifying a fault, is the first step towards ensuring the safety of a system. Once a fault is identified during the testing process, then \textit{debugging}, i.e. locating and repairing the fault, must be carried out to ensure safety and reliability of the system \cite{fault-local, repair-survey}. 
Although debugging is a distinct and equally critical stage of the development process, it has received significantly less attention than testing. In particular, no dedicated systematic review examines the techniques and challenges of debugging ADS. 
Additionally, the available debugging techniques are spread across various fields such as Software Engineering (SE) and Artificial Intelligence (AI), making it difficult to gain a comprehensive understanding of all the techniques.

To broaden our understanding of existing tools, techniques, and concepts related to debugging in ADS, we conducted a systematic review. There are two main types of systematic reviews: Systematic Literature Review (SLR) and Systematic Mapping Study (SMS). 
We chose to conduct an SMS due to the limited research on debugging in ADS. This approach allows us to provide a broad overview of the existing research in this area and map existing studies \cite{PETERSEN20151}, rather than narrowing our focus to specific research questions, which is typical of a systematic literature review \cite{BRERETON2007571}. Our aim was to explore the breadth, trends, and gaps in the literature, making the systematic mapping study a more appropriate choice.

This paper introduces the first SMS focused on ADS debugging. We adopt a structured methodology that includes strict inclusion and exclusion criteria along with a validation step, enabling us to categorise the existing literature and outline the current state-of-the-art (SOTA). Additionally, we propose a structured taxonomy based on the literature reviewed, along with classifications and terminology useful for future research in ADS debugging. The paper also contains a comprehensive analysis of trends and gaps within the ADS debugging literature, leading to a discussion on potential future research directions, which offer valuable insights for advancing the field.

Through this study, we provide guidance for ADS developers by listing tools to address various debugging challenges. Also, highlighting the importance of debugging in ADS will encourage future work to incorporate this stage explicitly, along with rigorous testing, utilising cross-disciplinary insights to improve associated methods. Our taxonomy and classifications in this field should establish a foundation for consistent terminology in future literature. Our work has the potential to aid in the increased reliability and safety of ADS, thereby accelerating development cycles and directing research to advance ADS towards commercial use.

The remainder of the paper is organised as follows. Section \ref{sec:background} provides a background on ADS, followed by related work in Section \ref{sec:related-work}. Section \ref{sec:methodology} presents our methodology for a systematic study. Sections \ref{sec:types} to \ref{sec:tools} present the results of our mapping, covering study types, debugging problems, techniques and tools. This is followed by an overarching discussion in Section \ref{sec:discussion} and the conclusion of our work in Section \ref{sec:conclusion}.

%% file: Sections/Background.tex
\section{Background}\label{sec:background}

This section presents background materials, including general ADS structure and conventional debugging of software systems. 

\subsection{ADS Structure}

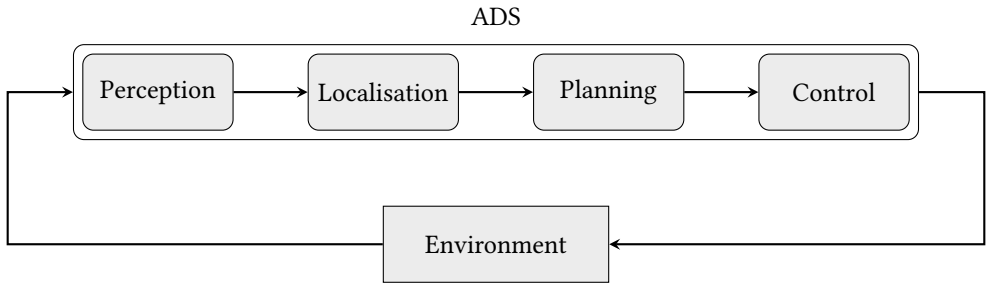
\begin{figure}
    \centering
    \scalebox{1.0}{
    \begin{tikzpicture}[node distance=2cm]
        \node (ads) [roundedbox] {
            \begin{tikzpicture}[node distance=3cm]
                \node (per) [smallbox] {Perception};
                \node (loc) [smallbox, right of=per] {Localisation};
                \node (plan) [smallbox, right of=loc] {Planning};
                \node (cont) [smallbox, right of=plan] {Control};
        
                \draw [arrow] (per) -- (loc);
                \draw [arrow] (loc) -- (plan);
                \draw [arrow] (plan) -- (cont);
            \end{tikzpicture}
        };

        \node [yshift=1cm] {ADS};

        \node (env) [process, below of=ads] {Environment};

        \draw [arrow] (env)  -- ++(-6.5,0) -- ++(0,2) --  (ads);
        \draw [arrow] (ads) -- ++(6.5,0) -- ++(0,-2) -- (env);
    \end{tikzpicture}
    }
    \caption{Basic module-based ADS Pipeline}
    \Description{Flow diagram detailing the four key modules of an ADS. Starting with Perception into Localisation, then Planning, followed by Control}
    \label{fig:pipeline}
\end{figure}

Broadly speaking, an ADS architecture can be classified into two types \cite{Yurtsever}: \textit{module-based} (i.e. composed of multiple, often heterogeneous modules) and \textit{end-to-end} (i.e. composed entirely of machine learning models). 
This study focuses on module-based ADS due to their greater explainability and practical relevance in industrial settings. 
These systems decompose functionality into basic modules (see \figurename~\ref{fig:pipeline}), enabling engineers to inspect, analyse, and debug each stage of the pipeline. 
In contrast, end-to-end systems function as deep learning black boxes, where debugging largely mirrors generic neural network analysis and offers limited insight into ADS-specific behaviour. 
Furthermore, many existing debugging techniques assume access to internal modules, an assumption only valid for module-based architectures. 

The four key modules in module-based ADS are as follows:
\begin{itemize}[-]
    \item \textit{Perception}: This module is responsible for gathering and combining all the input data from the environment. Input sources vary depending on the ADS but are generally comprised of camera and LIDAR data.
    \item \textit{Localisation}: This module determines the precise position of the vehicle within its environment. This uses the combined sensor data to locate the position of the ego vehicle in real time.
    \item \textit{Planning}: This module is responsible for taking the current position of the ADS from the localisation module and creating a path plan to a pre-determined destination. This generally does not involve planning the full route to the final destination, but a short-term plan to a sub-destination that can be executed as the next step.
    \item \textit{Control}: This module takes the plan created in the Planning module and transforms it into a set of instructions for the ego vehicle. This takes the form of speed and steering instructions, which the ego vehicle's hardware can execute. 
\end{itemize}

Beyond the complexity introduced by potentially heterogeneous modules, a key challenge in debugging ADS lies in the closed-loop interaction between the system and its environment. During a single scenario execution, individual modules are invoked repeatedly (often hundreds or thousands of times) with their outputs continuously influencing the following sensor inputs. This dynamic feedback loop significantly amplifies the difficulty of identifying the root cause of failures. 

\subsection{Conventional Debugging of Software Systems} \label{conventional-debug}

Understanding ADS debugging approaches requires knowledge of testing and debugging techniques used for conventional software systems. 
Debugging refers to localising the root causes of faults in systems, understanding their cause, and repairing incorrect behaviours \cite{Zeller2009, fault-local, repair-survey}. The debugging process is completed as a response to failing tests and serves as the next step towards the end of any development cycle. Debugging differs from testing, which is finding where a software program does not meet its specification. \citet{5386906} discusses the definition of testing as a combination of white-box methods applied during and after development, as well as black-box testing methods completed post-development to ensure the testing criterion is met. 
Methods such as Static Analysis or Delta Debugging \cite{Holger2005} can first be used to localise a software system's faulty code. Once the root cause of incorrect behaviour has been identified, a developer can begin to understand why the incorrect behaviour occurred and apply a repair technique to the software. Testing can be additionally used during the repair stage to verify if the fault has been fixed, as the original failing test should pass once the repair is successful. The importance of systematic debugging and its close ties to testing are covered extensively by \citet{Zeller2009}, in which many more aspects and methods of debugging are discussed.

In conventional software systems, the debugging process is commonly structured into three interrelated tasks: fault \textit{localisation}, fault \textit{explanation}, and fault \textit{repair}.
Localisation refers to the task of identifying the precise location of the fault within the program, such as erroneous lines of source code or malfunctioning components.
Explanation involves understanding the underlying mechanism by which a given input leads to the observed failure, often requiring an analysis of program state and control flow.
Repair is concerned with modifying the faulty code to eliminate the failure, ideally without introducing unintended side effects.

In the following subsections, we describe each of these tasks in more detail, along with representative techniques and tools that support them.

\subsubsection{Localise}
Fault localisation is typically a first step in conventional debugging approaches with the goal of identifying code locations likely to contain faults. \citet{fault-local} surveyed fault localisation techniques extensively, covering slice-, spectrum-, state-based and more. 

Spectrum-based fault localisation (SBFL) \cite{fault-local} is a statistical approach used to rank program entities based on their likelihood of being faulty. It analyses a program's execution spectra, i.e. parts of code executed by passing and failing tests. Code can be grouped into sections based on execution data, with various ranking metrics used to determine how suspicious each section is. SBFL is known for its scalability for larger software and simple implementation; however, it has its drawbacks. One limiting factor is in domains where correctness is harder to define (e.g. autonomous systems) because individual modules (and by extension their functions and lines of code) are invoked multiple times within a single scenario execution. This repeated invocation complicates the identification of passing and failing behaviours at the invocation level, making traditional notions of test oracles less straightforward to apply \pref{P10}.

Delta debugging, covered in \citet{Holger2005}, is another well-known approach for minimising test inputs and localising faults. It works by repeating a failed test with different aspects of the test inputs removed, aiming to preserve faulty behaviour. By doing this, a simplified test case is presented that only contains key elements/components responsible for the failure. This is a common approach for debugging methods and has direct or indirect application in other studies. \citet{Zeller2009} highlights ``cause-effect chains'', underscoring the importance of fault localisation, as each failure can be traced back to a root cause in the code. 

Another widely used approach to fault localisation, algorithmic debugging, is explored extensively in \citet{Algorithmic_debugging_survey}. In general terms, this method functions by creating execution trees of program computations such as function calls. A developer is then queried on each step of the tree, checking the results of each logic step in a language-independent form. By pruning the tree, errors can be traced to their associated area of the code. One drawback of this approach is the scalability, with a system designed for autonomous driving likely producing a tree that is too large to be assessed by individuals. Setting results within the tree also assumes reproducibility for each function, which may not always be the case with cyber-physical systems (CPSs) such as ADS, as external nondeterministic factors can affect the outcome of the scenario in real-world situations. Methods discussed in \cite{Algorithmic_debugging_survey} could potentially be applied to ADS, showing potential for human-in-the-loop debugging methods, which can be reworked or scaled for a more complex system.

\subsubsection{Explain}
Explanation is another key step to debugging a system, as understanding the nature of a system and its faulty behaviour can facilitate repair. This aspect of debugging usually follows localisation in conventional software systems and has a large focus in machine learning (ML) settings, likely due to the explanation of ML models being more difficult. As highlighted by \citet{castelvecchi_can_2016}, many ML applications act as black box systems, so understanding the problem is a much more difficult task. The field of Explainable AI (XAI) aims to tackle this with works looking to explain ADS behaviours \cite{XAI-survey}, where explanations of failures overlap with debugging.

One notable explanation approach is Shapley Additive Explanations (SHAP), introduced by \citet{SHAP2017}, used to explain the actions of a machine learning model. This method assigns importance values to input features based on their contribution to a model's output. By assigning weights to input features, SHAP creates interpretable links between the model's inputs and outputs, making it easier to identify why an input may have led to a model having a particular output. Regarding debugging, this allows a developer to identify what input contributed to a system failure and gain insights into potentially faulty system components.

Despite appearing as a step mostly applicable to ML, there are explainable techniques that can be applied to standard software systems. Explanations of software can take the form of sequence diagrams visualising the steps involved in a system to help developers understand. These approaches are covered in detail in the survey by \citet{program_comprehension}, which contains common approaches to explain software structure and behaviour. A major goal of such techniques is to explain faults through trace visualisation or feature location, assisting the debugging process by enhancing developers' understanding of the system as a whole. Due to the large volume of systematic literature on explainable AI we chose to solely focus on the explanations of failures within this study.

\subsubsection{Repair}
In conventional debugging, repair follows fault localisation and may involve either manual code revision or automated techniques designed to accelerate debugging, especially in large-scale systems. Researchers have explored automated software repair to reduce reliance on manual debugging efforts. \citet{Automatic_software_repair_survey} explores methods of repair and distinguishes types, proposing the definitions of software healing and repair as two distinct tasks. Software healing refers to correcting system states in runtime to return a system to correct operational status. In contrast, software repair is the automated generation of new code to correct faulty behaviour in all future code executions. Healing techniques are more applicable to fault tolerance, which is beyond the scope of this paper; however, repair techniques show promise in the field of ADS. \citet{Automatic_software_repair_survey} proposes many search-based and learning-based techniques with heavy reliance on data-driven methods and evolutionary algorithms to produce software patches. These techniques are particularly promising in large-scale systems, where manual repair by individual developers would be time-consuming and less scalable. Many of these repair techniques rely on existing fault localisation approaches, highlighting their role as a next step process.

Typical repair methods involve finding alternative code to replace the fault-triggering code in the system. This new code must maintain correct outputs for the system whilst also removing the fault located previously. One common way to find alternative code or system behaviours is to use Genetic Algorithms (GA) \cite{genetic-algorithm} or similar meta-heuristics. By first providing a starting point, a genetic algorithm can be used to search for code alternatives automatically. A challenge to this approach, as discussed by \citet{fault-local}, is to avoid over-fitting, where GA-generated code becomes too specialised to the test suite, causing new faults in other scenarios. The threat of over-fitting also applies to other automated repair approaches, such as with data-driven ones, so any future implementation needs specific constraints to generate successful code patches.

As with explanation, automated repair approaches are prevalent in the field of ML due to the complexity of manual repair. Methods to automatically correct the outputs of a model are gaining popularity, such as the method introduced by \citet{Arachne}. DNN (deep neural network) patching can be used to target specific neurons within a network which are responsible for incorrect output. These can then be patched to give correct output without completely retraining the model. For the purpose of this study, we only consider model repair methods specifically applied to ADS.

Reliability of automated repair techniques is also a concern in the field, with one notable study by \citet{Automated_debugging_helping} which investigates if automated tools are truly benefiting development. They found that ranking-based tools often fall short in practice. Tools automatically suggest potentially faulty code to assist developers, but they do not always reduce repair time, as their approach often misaligns with real-world debugging practices. This study makes a call for more human-like debugging tools that will help developers understand sources of code faults. Automated techniques have developed since this review and show potential in large-scale systems such as ADS, with many ADS developers already implementing automated tools into practice.

\section{Related Work}\label{sec:related-work}

Given the close relationship between ADS testing and debugging, understanding the current testing landscape is related to our mapping study. 

Several systematic literature surveys provide an overview of ADS testing \cite{Araujo, Lou2022, Zahra2022, Tang2023}. 
\citet{Araujo} offers a comprehensive overview of testing robotic and autonomous systems, including ADS, while also covering validation and verification (V\&V) practices, providing valuable insights across domains. \citet{Lou2022} gives a similar overview with a specific interest in industrial practices and academic techniques, explaining many relevant ADS testing techniques. Despite their comprehensive scope, both surveys limit their focus to failure identification, without addressing localisation or repair. This trend continues within other notable testing surveys such as \citet{Zahra2022}, which details test case generation, selection, and prioritisation. This study is concerned with improving the effectiveness and efficiency of test processes without considering the post-failure activities involved in debugging. 
\citet{Tang2023} introduces different approaches to testing for module-level versus system-level ADS, an important distinction for debugging since the level at which a failure manifests influences the techniques to localise and resolve. 

Scenario-based testing is the testing of ADS in real world scenarios, often through simulation. This method is a foundational strategy in ADS development and shares natural ties with scenario-based debugging, though this connection remains under-explored in the literature. \citet{Ding2023} and \citet{Schütt2023} both provide comprehensive taxonomies of scenario generation for ADS testing. These surveys emphasise the importance of formalised scenario generation techniques to meet safety and regulatory standards. \citet{Sun2022} further narrows the focus, looking specifically into test automation for ADS, categorising strategies by their coverage capabilities. Another more specialised survey is presented by \citet{zhong2021}, which focuses on techniques specifically tailored for high-fidelity simulated environments. The survey identifies challenges in evaluating critical scenarios and highlights the growing use of software-in-the-loop testing as a scalable and safer alternative. Finally, \citet{Zhang2023} presents an SMS on critical scenarios for ADS, emphasising the importance of scenario-based testing in improving safety, identifying unknown hazards, and refining operational design domains. 

Unlike other simulation surveys, the work of \citet{Corso_2021} thoroughly examines black-box safety validation algorithms for cyber-physical systems, including ADS. It categorises SOTA techniques from optimisation, path planning, reinforcement learning, and sampling, all aimed at identifying safety violations in simulated environments. 
Another forward-looking approach is offered by \citet{birchler2024}, which outlines challenges in simulation-based testing for autonomous cyber-physical systems (ACPSs), with a focus on issues such as test case formulation, simulation realism, the oracle problem, and cost-effective regression testing strategies. 

Across all these systematic reviews, while they offer valuable foundations for understanding ADS testing, the emphasis remains overwhelmingly on testing itself, with limited attention given to post-failure analysis. To date, there is no systematic review dedicated to ADS debugging, leaving a significant research gap. Our systematic mapping study aims to address that gap by examining the research types, problem spaces, techniques, and tools that characterise the current landscape of debugging in ADS.

%% file: Sections/Methodology.tex
\section{Methodology}\label{sec:methodology}
\begin{figure}
    \centering
    \begin{tikzpicture}[node distance = 3cm]
        \node (plan) [process] {Planning the review};
        \node (con) [process, below of=plan] {Conducting the review};
        \node (rep) [process, below of=con] {Reporting the review};
        
        \node (p) [roundedbox, right of=plan, xshift=3cm] {Research Questions \& Scope (\S~\ref{sec:methodology:planning})};
        
        \node (c) [roundedbox, right of=con, xshift=3cm] {
            \begin{tikzpicture}[node distance=1.1cm, every node/.style={font=\scriptsize}]
                    \node (s1) [process, minimum width=2.5cm, minimum height=0.8cm] {Seed Paper Selection};
                    \node (s2) [process, below of=s1, minimum width=2.5cm, minimum height=0.8cm] {Literature Search \& Filtering};
                    \node (s3) [process, below of=s2, minimum width=2.5cm, minimum height=0.8cm] {Data Analysis};
                    \draw[arrow] (s1) -- (s2);
                    \draw[arrow] (s2) -- (s3);
                \end{tikzpicture}
            (\S~\ref{sec:methodology:conducting})
        };
        
        \node (r) [roundedbox, right of=rep, xshift=3cm] {Results \& Discussions (\S~\ref{sec:types}-\ref{sec:discussion})};

        \draw[arrow] (plan) -- (con);
        \draw[arrow] (con) -- (rep);
        \draw[dottedarrow] (plan) -- (p);
        \draw[dottedarrow] (con) -- (c);
        \draw[dottedarrow] (rep) -- (r);
    \end{tikzpicture}
    \caption{Steps of the mapping study adapted from \citet{Zhang2023}}
    \Description{Flow diagram of the mapping study steps. Starting with planning the review which contains defining research questions and scope. This is followed by conducting the review which contains a sub process diagram of seed paper select, to literature search and filtering, ending with data analysis. This chart ends with Reporting the review which is this paper}
    \label{fig:steps}
\end{figure}
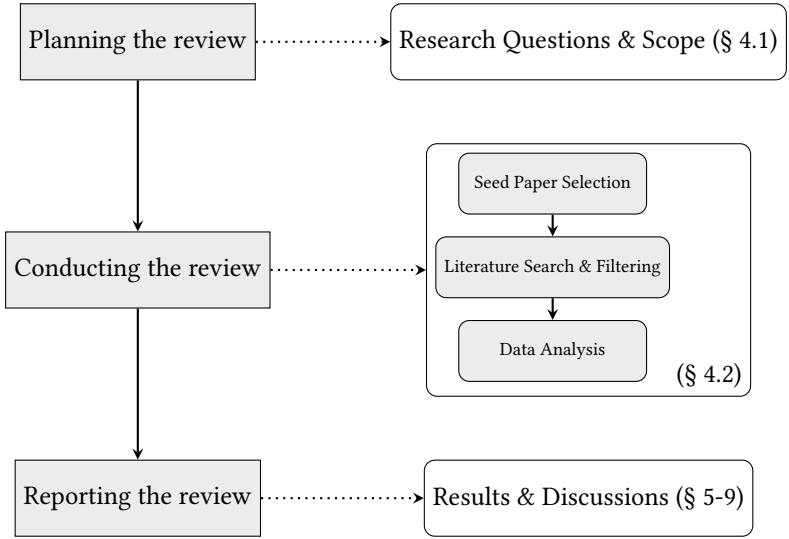

This mapping study adheres to the review steps and guidelines set out by \citet{PETERSEN20151}, and draws structural inspiration from \citet{Zhang2023}, which informs the division of the study into three distinct stages as illustrated in \figurename~\ref{fig:steps}: (1) planning, (2) conducting, and (3) reporting.

\subsection{Planning the Review}\label{sec:methodology:planning}
We initiated the planning of the SMS with the following research questions:
\begin{enumerate}[\bf RQ1]
    \item What \textit{types of research} have been conducted on debugging in ADS?
    \item What specific \textit{problems} within ADS can be identified and addressed through debugging?
    \item What debugging \textit{techniques} have been utilised in current ADS debugging studies?
    \item What existing debugging \textit{tools} are currently available for ADS?
\end{enumerate}

RQ1 provides an overview of the field by highlighting research types, helping to identify trends and under-explored areas. 
Similarly, RQ2 identifies the types of problems addressed in ADS debugging papers. This is required to understand the actual types of problems that exist in ADS and what has been done to address them. Identifying problem types and how they are addressed can also reveal gaps and give specific targets for future research. 
RQ3 finds what techniques have been used in the field. Provide a specific list of applicable techniques that can show what conventional debugging techniques are applicable to ADS, as well as suggest methods that have not yet been applied. In addition to techniques, tools also play a crucial role. 
RQ4 finds what tools have been used and what tools have been purposely built for debugging in ADS. Findings from RQ4 should help researchers select appropriate tools to tackle specific problems or identify opportunities to develop new tools based on the problem type.

To facilitate a more focused review, we employed the PICOC framework (Population, Intervention, Comparison, Outcomes, and Context) to formulate the research questions and define the inclusion and exclusion criteria, in accordance with the guidelines of \citet{keele2007guidelines}. Specifically, the scope of our review is characterised as follows:
\begin{description}
    \item[Population] Autonomous Driving Systems (ADS)
    \item[Intervention] Debugging techniques applied to ADS
    \item[Comparison] Not applicable
    \item[Outcomes] Improved software performance
    \item[Context] Peer-reviewed publications
\end{description}

By following these guidelines, we established a clearer understanding of the types of studies relevant to our investigation. In particular, we focused on works that propose methods to actively address or act on faults, rather than merely identifying them. We chose not to impose a minimum citation count, as the research area is relatively new and emerging. Based on our domain knowledge and research questions, we developed a set of inclusion and exclusion criteria to guide the literature search.

\paragraph{Inclusion Criteria}
\begin{enumerate}[\bf I1]
    \renewcommand{\theenumi}{\bfseries I\arabic{enumi}:}
    \renewcommand{\labelenumi}{\theenumi}
    \item The study must focus on on-road autonomous (automated) driving.
    \item The study must address "debugging," rather than "testing."
\end{enumerate}

\paragraph{Exclusion Criteria}
\begin{enumerate}[\bf E1:]
    \item Studies published before January 2015 or after February 2025 (since the review was conducted between February-March 2025). 
    \item Studies not written in English.
    \item Non-primary studies (e.g. survey, mapping studies).
    \item No duplicate papers.
\end{enumerate}

The exclusion criteria were formed to reduce search results automatically as applied filters. By filtering studies older than ten years, we only capture the current, more relevant state of the field. 
Only including studies written in English was set, so there were no issues during the filtering and data extraction stages. As this mapping study constitutes secondary research, our search exclusively targets primary studies that present original contributions.

In addition to these criteria, we explicitly exclude general research on Explainable AI (XAI), which has already been extensively studied and systematically reviewed, unless it directly addresses ADS failures—for example, \citet{XAI-paper}. Specifically, studies that explain general model decisions are excluded, whereas studies that aim to explain why a model made an incorrect or unsafe decision are included. Similarly, studies that focus on connected autonomous vehicles (CAV), particularly where networking aspects dominate, like in \citet{CAV-paper}, are considered out of scope. These areas are distinct research domains with objectives beyond the scope of this study.

Our inclusion criteria are that the study must focus on debugging in ADS. Unlike exclusion criteria, they cannot be applied automatically and require manual interpretation of the material. Once the filtered results list was returned from the exclusion criteria filters, we proceeded to manually review the papers to ensure they focused on the correct topic. 

\subsection{Conducting the Review} \label{sec:methodology:conducting}

\begin{figure}
    \centering
    \scalebox{0.75}{
    \begin{tikzpicture}[node distance=3.5cm, yshift=-0.5cm]

        \node (step1) [process] {Seed Paper Collection};
        \node (step2) [process, right of=step1, xshift=4cm] {Create Search String};
        \node (step3) [process, below of=step1] {\parbox{4cm}{Collect Search Results \\ with Exclusion Criteria}};
        \node (step4) [process, right of=step3, xshift=4cm] {\parbox{4.5cm}{\centering Paper Filtering \\ (Title, TAK, Deep Filtering)}};
        \node (step5) [process, below of=step3] {\parbox{4.5cm}{\centering Discussion \\ and Conflict Resolution}};
        \node (step6) [process, right of=step5, xshift=4cm] {Snowballing};
        \node (decision) [decision, below of=step6] {Have the Stopping Criteria Been Met?};
        \node (step7) [process, below of=decision] {Data Extraction};

        \draw [arrow] (step1) -- node[midway, above] {10 Seed Papers} (step2);
        \draw [arrow] (step2) -- (step3);
        \draw [arrow] (step3) -- node[midway, above] {2,000 Papers} (step4);
        \draw [arrow] (step4) -- node[midway, left] {R1:43, R2:30} (step5);
        \draw [arrow] (step5) -- node[midway, above] {R1:14, R2:15} (step6);
        \draw [arrow] (step6) -- node[midway, right] {R1:15, R2:15} (decision);
        \draw [arrow] (decision) -- node[midway, left] {Yes} node[midway, right] {15 Final Papers} (step7);
        \draw [arrow] (decision.east) -- ++(1,0) |- node[midway, right, yshift=-5cm] {No} (step2);

    \end{tikzpicture}
    }
    \caption{Overview of the Systematic Search and Selection Process (R1: Round 1, R2: Round 2 using the Snowballing Results from R1)}
    \Description{Flowchart of the research process starting with Seed paper collection, followed by creating a search string. This is followed by collecting search results with applied exclusion criteria into applying the inclusion criteria. A manual reading step follows with a discussion step afterwards, which feeds back into the reading step. Backward Snowballing is next which leads to a decision to repeat. If yes go back to the search string step if no move to data extraction}
    \label{fig:research-cycle}
\end{figure}
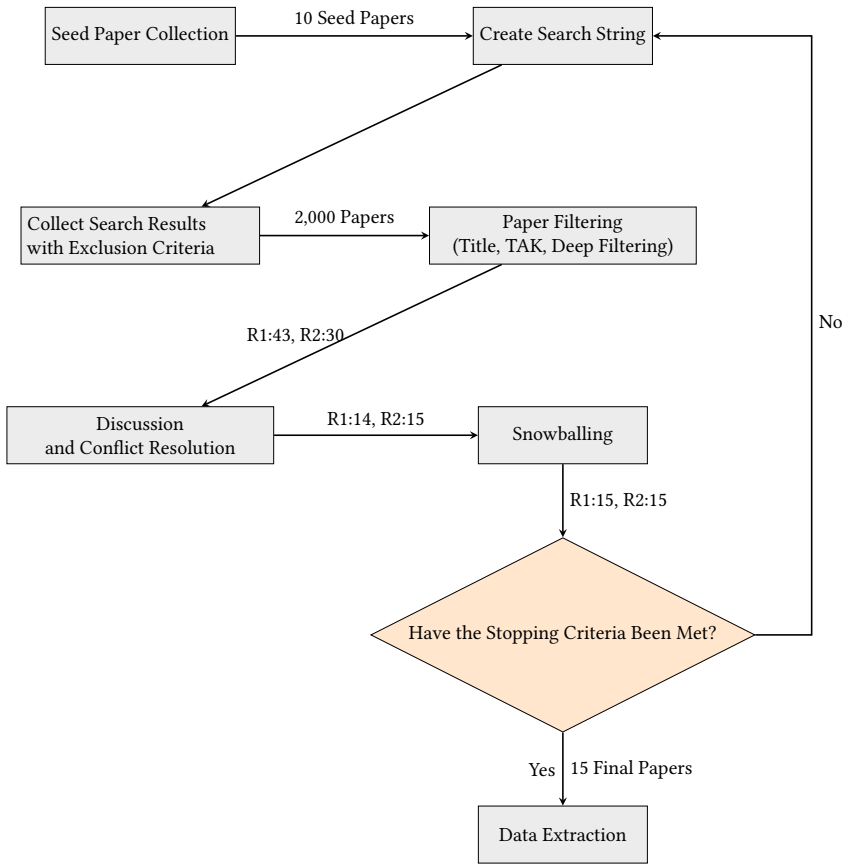

Collecting relevant papers for this study required a rigorous search and filtering process detailed in Figure \ref{fig:research-cycle}. 
This process begins with seed paper collection, including relevant papers used to generate key terminology to formulate a search string. 
The search string retrieves a list of publications from a database search, which is filtered using the previously defined exclusion criteria. 
The inclusion criteria are applied manually to the remaining results. 
In cases of disagreement or uncertainty, papers are discussed among reviewers in a recursive decision process. 
Discussing conflicts between reviewers at this stage ensures that only relevant papers are included in the final shortlist. 
Following this, snowballing \cite{Wohlin2014} is utilised to find other related papers and check if any key search terms have been missed. 
If new key terms are found, the process is repeated with a new search string updated with the new key terms. The process continues until backward snowballing no longer reveals new relevant keywords. At that point, the review proceeds to the data extraction phase

\subsubsection{Seed Paper Collection}
To start a systematic mapping study, we first need a set of starting papers relevant to the field, which can be used as a base for our search. These papers help define the scope of the field and are used to extract relevant keywords. A database search can then be conducted using the key terms collected (see Section \ref{sec:database-search}). 

Initially, a small set of ADS debugging papers was obtained through domain knowledge from one of the authors. We identified key terms from these papers and became familiar with the relevant terminology. Because the Comparison, Outcome, and Context elements from the PICOC framework were still under development at this stage, we restricted our search on Google Scholar to terms related to Population and Intervention as previously discussed. This process expanded our list from 3 to 22 candidate papers. We then reviewed these papers by examining their Titles, Abstracts, and Keywords (TAK) \cite{BRERETON2007571}. TAK filtering enables rapid screening of papers by assessing only their TAK, without requiring a full-text review. This faster process is essential when screening hundreds of papers, as needed for this study. Relevant papers were recorded in a shared spreadsheet \cite{SMSdataset}, facilitating documentation, tracking, and later data extraction. Papers that were not directly related to ADS debugging were excluded, leaving us with 10 seed papers.

\subsubsection{Database Search}\label{sec:database-search}
We extracted keywords from the 10 seed papers based on their title, keywords, and abstracts and formed our initial search string. Table \ref{tab:search-string-final} shows the search string (except for the bold ones, which were added after snowballing; see Section~\ref{sec:snowballing}).
To ensure comprehensive coverage of relevant literature, we structured the search string into two components: terms related to ADS and terms related to debugging. The final search string was formed by combining these two components using a logical AND conjunction.

\begin{table}
    \centering
    \small
    \caption{Search String (the keywords added after the initial search string are highlighted in \textbf{bold})}
    \begin{tabularx}{\linewidth}{lX}
        \toprule
        \textbf{Term Type} & \textbf{Search String} \\
        \midrule
        ADS Term &
        ("ADS" OR "Automated driving" OR "Autonomous Driving" OR "Autonomous Vehicle" OR "Self-driving Car") \\
        Debugging Term &
        ("debug" OR "diagnose" OR "diagnosis" OR "explanation" OR "explain" OR "localise" OR "localize" OR "localisation" OR "localization" OR "isolate" OR "minimise" OR "minimize" OR "minimisation" OR "minimization" OR "simplification" OR "simplify" OR "causal" OR "repair" OR "\textbf{root cause}") \\
        \bottomrule
    \end{tabularx}
    \label{tab:search-string-final}
\end{table}

We selected OpenAlex \cite{Priem2022OpenAlex} as the primary database for our systematic mapping study due to its comprehensive and up-to-date coverage of scholarly literature. OpenAlex indexes over 250 million works across disciplines, including publications from major sources such as IEEE and ACM, and has been demonstrated to subsume the contents of Scopus \cite{Alperin2024AnAO}. Furthermore, its support for metadata filtering enabled us to apply exclusion criteria automatically, improving the rigour and reproducibility of our selection process.
As a result of executing the initial search string on OpenAlex, combined with automated filtering based on our exclusion criteria, we retrieved a total of 6,641 papers. 

Given the impracticality of manually screening all 6,641 papers based on our inclusion criteria, we leveraged the relevance ranking of OpenAlex to narrow the list. We found that all 10 seed papers appeared within the top 1,300 results, confirming the relevance score's effectiveness. To remain conservative while ensuring a manageable review workload, we selected the top 2,000 most relevant results for manual screening against the inclusion criteria.

\subsubsection{Paper Filtering}\label{sec:paper-filtering}
From the 2,000 most relevant search results, we manually applied the inclusion criteria through a four-step filtering process: (1) title filtering, (2) TAK filtering, (3) deep filtering, and (4) conflict resolution.

The first step was title filtering, which aimed to narrow the list to potentially relevant papers based on titles alone. The first two authors independently assessed each title against the inclusion criteria (I1 and I2), assigning one of three labels: Yes, No, or Unsure. A paper was excluded only if both reviewers marked it as No. This step resulted in 117 papers selected for further review.
The second step was TAK filtering, which followed the same procedure as title filtering, but reviewers used titles, abstracts, and keywords to inform their decisions. 
The third step, deep filtering, required reviewers to read the full texts of the remaining papers and apply the same review protocol. After applying all of the inclusion criteria we were left with 43 papers in round 1 and 30 in round 2.
For papers excluded in the title, TAK, and deep filtering steps, the third and fourth authors conducted an additional validation to identify any papers that might warrant further discussion. Such papers were then moved to the final, conflict resolution step. Over the two round of studies 10 papers were brought back into discussions via validation steps.
In the final step, all four authors discussed any remaining disagreements. Through the conflict resolution process, 14 papers were selected for snowballing in the first round (R1). One additional paper was selected in the second round (R2), following a repetition of the entire process in Figure \ref{fig:research-cycle} using the updated search string after snowballing.
All decisions made at each step for every paper are documented in the supplementary material \cite{SMSdataset}.

\subsubsection{Snowballing}\label{sec:snowballing}
Following established guidelines for systematic mapping studies \cite{PETERSEN20151}, we applied backward snowballing \cite{Wohlin2014} to the 14 papers identified in the first round (15 in the second round), reviewing their reference lists to uncover additional relevant literature. This approach helps reduce the risk of overlooking important studies not captured in the initial database search. We did not apply forward snowballing, which identifies newer papers citing the shortlisted ones, as such studies were expected to be included in the initial search results.

The snowballing process was primarily to identify any new keywords to refine or expand the search string. Following the first round of snowballing, we identified a new keyword (Root Cause), which was added to the search string before restarting the process in Figure \ref{fig:research-cycle}. No new keywords were found after the snowballing in the second round with the refined search string. This indicated that conceptual coverage had been achieved, and further snowballing was unlikely to contribute additional value. As a result, the search was concluded, yielding 15 final papers for data extraction. 

For clarity and ease of reference, all the 15 papers included after completing the snowballing process were assigned unique Paper IDs (P1–P15) as shown in Table \ref{tab:paperids}. These identifiers are used throughout the remainder of the paper when referring to individual studies. 

\begin{table}
    \centering
    \rotatebox{90}{
    \begin{minipage}{\textheight}
    \caption{Overview of Included Papers (P1–P15)}
    \label{tab:paperids}
    \scriptsize
    \renewcommand{\arraystretch}{1.5}
    \rowcolors{2}{gray!10}{white}
    \begin{tabular}{p{0.8cm} p{6.5cm} p{7cm} p{3cm} p{0.5cm}}
        \toprule
        \textbf{Paper ID} & \textbf{Title} & \textbf{Authors} & \textbf{Venue} & \textbf{Year} \\
        \midrule
\plabel{P1} & Debugging Autonomous Driving Systems Using Serialized Software Components \cite{paper7} & Pascal Minnerup, David Lenz, Tobias Kessler, and Alois Knoll. & IFAC-PapersOnLine & 2016 \\
\plabel{P2} & Autonomous Driving—A Crash Explained in Detail \cite{paper1} & Johannes Betz, Alexander Heilmeier, Alexander Wischnewski, Tim Stahl, and Markus Lienkamp. & Applied Sciences & 2019 \\
\plabel{P3} & Automated Repair of Feature Interaction Failures in Automated Driving Systems \cite{paper4} & Raja Ben Abdessalem, Annibale Panichella, Shiva Nejati, Lionel C. Briand, and Thomas Stifter. & ISSTA & 2020 \\
\plabel{P4} & A Comprehensive Study of Autonomous Vehicle Bugs \cite{paper5} & Joshua Garcia, Yang Feng, Junjie Shen, Sumaya Almanee, Yuan Xia, and Qi Alfred Chen. & ICSE & 2020 \\
\plabel{P5} & A Fault Detection and Diagnosis System for Autonomous Vehicles Based on Hybrid Approaches \cite{paper2} & Yukun Fang, Haigen Min, Wuqi Wang, Zhigang Xu, and Xiangmo Zhao.  & IEEE Sensors Journal & 2020 \\
\plabel{P6} & Parameter-Based Testing and Debugging of Autonomous Driving Systems \cite{paper6} & Paolo Arcaini, Alessandro Calò, Fuyuki Ishikawa, Thomas Laurent, Xiao-Yi Zhang, Shaukat Ali, Florian Hauer, and
Anthony Ventresque. & IEEE IV Workshops & 2021 \\
\plabel{P7} & Less is More: Simplification of Test Scenarios for Autonomous Driving System Testing \cite{paper16} & Paolo Arcaini, Xiao-Yi Zhang, and Fuyuki Ishikawa. 2 & ICST & 2022 \\
\plabel{P8} & Fault Diagnosis of the Autonomous Driving Perception System Based on Information Fusion \cite{paper18} & Wenkui Hou, Wanyu Li, and Pengyu Li. & Sensors & 2023 \\
\plabel{P9} & Explaining a Machine-Learning Lane Change Model With Maximum Entropy Shapley Values \cite{paper8} & Meng Li, Yulei Wang, Hengyang Sun, Zhihao Cui, Yanjun Huang, and Hong Chen.  & IEEE T-IV & 2023 \\
\plabel{P10} & Towards Automated Driving Violation Cause Analysis in Scenario-Based Testing for ADS \cite{paper20} & Ziwen Wan, Yuqi Huai, Yuntianyi Chen, Joshua Garcia, and Qi Alfred Chen.  & arXiv & 2024 \\
\plabel{P11} & ROCAS: Root Cause Analysis of Autonomous Driving Accidents via Cyber-Physical Co-Mutation \cite{paper9} & Shiwei Feng, Yapeng Ye, Qingkai Shi, Zhiyuan Cheng, Xiangzhe Xu, Siyuan Cheng, Hongjun Choi, and Xiangyu Zhang. & ASE & 2024 \\
\plabel{P12} & ADAssure: Debugging Methodology for Autonomous Driving Control Algorithms \cite{paper14} & Andrew Roberts, Mohammad Reza Heidari Iman, Mauro Bellone, Tara Ghasempouri, Jaan Raik, Olaf Maennel,
Mohammad Hamad, and Sebastian Steinhorst. & DATE & 2024 \\
\plabel{P13} & ACAV: Automatic Causality Analysis in Autonomous Vehicle Accident Recordings \cite{paper15} & Huijia Sun, Christopher M. Poskitt, Yang Sun, Jun Sun, and Yuqi Chen. & ICSE & 2024 \\
\plabel{P14} & Modular Fault Diagnosis Framework for Complex Autonomous Driving Systems \cite{paper3} & Stefan Orf, Sven Ochs, Jens Doll, Albert Schotschneider, Marc Heinrich, Marc René Zofka, and J. Marius Zöllner. & ICCP & 2024 \\
\plabel{P15} & A Comprehensive Study of Bug-Fix Patterns in Autonomous Driving Systems \cite{paper22} & Yuntianyi Chen, Yuqi Huai, Yirui He, Shilong Li, Changnam Hong, Qi Alfred Chen, and Joshua Garcia. & ACM Proc. Softw. Eng. & 2025 \\
        \bottomrule
    \end{tabular}
    \end{minipage}
    }
\end{table}

\subsubsection{Data Extraction} \label{Data_extraction}

Data extraction is the process of recording findings relevant to our research questions from each relevant paper. For this process, we again referred to the guidelines paper \cite{PETERSEN20151} and followed a similar approach. This process was partially conducted alongside deep filtering and snowballing, during which reviewers read each paper in full. Each paper used a form to record key details, including the title for identification and relevant information for each research question.

We ensure data validity by having a reviewer (first author) create and initial data extraction form, followed by a second reviewer (second author) checking this for each paper independently. The second reviewer used the form as an initial reference, then verified each recorded point by locating supporting evidence in the paper. This two-step approach ensures the validity and accuracy of our data extraction. 
Detailed results of the data extraction process are available in the supplementary material \cite{SMSdataset}. 
A more in-depth discussion of potential threats to the validity of our findings is provided in Section~\ref{threats}.

\subsection{Reporting}
Reporting was the final stage of this study and is presented in this paper (specifically, Sections \ref{sec:types} to \ref{sec:tools}). The results from data extraction are analysed and discussed, highlighting trends across studies and identifying gaps in the current knowledge. These insights provide a clear direction for future research in ADS debugging.

\subsection{Threats to Validity} \label{threats}

This study has potential threats to the validity of its results. This section highlights the main threats and discusses efforts to mitigate them.

\paragraph{Database Selection}
We identified that selecting OpenAlex is a source of potential bias, as there may be more papers available from other notable databases. 
To mitigate concerns, we conducted searches on other notable databases such as IEEE and ACM, confirming that all their relevant results were included within the OpenAlex results. OpenAlex's nature as a superset of Scopus \cite{Alperin2024AnAO} made it suitable for this study, ensuring coverage of a wide variety of databases and minimising the chance of missing relevant literature.

\paragraph{Paper Selection}
While manually applying the inclusion criteria, we found that one author's interpretation of a paper could differ from another's. Such differences introduce the risk of bias and may affect classification accuracy. To mitigate this, two authors independently conducted the initial filtering on the results using the titles, followed by the rest of TAK (title, abstract, keywords) as detailed in Section \ref{sec:methodology:conducting}. Having two authors involved in each stage reduced bias and facilitated discussions regarding paper inclusion. The other two authors then validated the excluded papers to ensure no papers were incorrectly filtered out. Flagged papers were reconsidered in discussions on the inclusion criteria and could potentially be reinstated into the relevant papers list. This rigorous process with multiple discussion stages ensured papers included were only those that passed the strict criterion of an ADS debugging paper.

\paragraph{Data Extraction}
During data extraction, bias could occur if a single author collected data, as their interpretation might differ from the actual content. To mitigate this, we used the two-tiered data collection process detailed in Section \ref{Data_extraction}. Once an extraction form is completed, a different author reviews the form whilst reading the source paper. Extracted data were verified against the paper, with any discrepancies discussed and corrected. Once the first two authors had finished this process, the form moved into a validation stage. During this stage, the remaining two authors completed similar steps to the previous validation, leaving comments for amendments to ensure data accuracy. Throughout discussions of data classifications and trends, the authors collaboratively made all data representation decisions.

%% file: Sections/Research_Types.tex
\section{RQ1: ADS Debugging Research Types}\label{sec:types}

\subsection{Taxonomy of Research Types}
We developed a taxonomy to categorise the different research types identified in the final 15 relevant papers. We classified them into three categories: \textit{validation}, \textit{evaluation}, and \textit{empirical} research, inspired by the software engineering research taxonomy provided by \citet{PETERSEN20151}. They are defined as follows:
\begin{itemize}[-]
    \item \textit{Validation Research}: Conducted in a controlled, academic or laboratory setting to demonstrate the feasibility of a technique, method, or approach. Such studies typically do not involve deployment in an industrial setting but instead focus on early-stage testing, simulation or proof-of-concept experiments using synthetic datasets.

    \item \textit{Evaluation Research}: Conducted in real-world or industrial contexts to address how a technique, method, or approach performs in practice. These studies involve real-world datasets, stakeholders, or operational environments, with the aim of understanding practicality and limitations in real-world settings.

    \item \textit{Empirical Research}: Observational study analysing existing data artefacts (e.g. ADS repositories and benchmark datasets) to derive insights without running a specific ADS.
\end{itemize}

For validation and evaluation types, we further classified their experimental environments to better understand the trends in ADS debugging studies. Specifically, we found the following experimental environment types:
\begin{itemize}[-]
    \item \textit{Offline Dataset}: Using a pre-collected labelled dataset, either synthetic or real-world, to investigate the ADS. The key characteristic is the lack of a feedback loop between the ADS and its operational environment, meaning the ADS's behaviour at any given time step does not affect its observation at the subsequent time step.
    \item \textit{Online Environment}: Conducted within a closed-loop manner, either simulation or real-world. The ADS is embedded and dynamically interacts with its environment, allowing for a feedback loop between the system and its surroundings.
\end{itemize}

\begin{figure}[ht]
    \centering
    \begin{tikzpicture}[node distance=1.5cm]
        \node (type) [process] {Research Type};
        
        \node (val) [process, right of=type, xshift=2cm] {Validation Research};
        \node (eval) [process, below of=val] {Evaluation Research};
        \node (emp) [process, below of=eval] {\parbox{3cm}{\centering Empirical Research \\ \pref{P4}, \pref{P15}}};

        \node (syn) [process, right of=val, xshift=3.5cm, yshift=1cm] {\parbox{5cm}{\centering Offline Synthetic Dataset-Based \\ \pref{P9}, \pref{P12}}};
        \node (sim) [process, below of=syn, yshift=-0cm] {\parbox{5cm}{\centering Online Simulation-based \\ \pref{P3}, \pref{P7}, \pref{P8}, \pref{P11}, \pref{P13}}};
        \node (off) [process, below of=sim, yshift=0cm] {\parbox{5cm}{\centering Offline Real-world Dataset-Based \\ \pref{P2}, \pref{P10}}};
        \node (real) [process, below of=off, yshift=0cm] {\parbox{5cm}{\centering Online Real-world Deployment-based \\ \pref{P1}, \pref{P5}, \pref{P6}, \pref{P14}}};

        \draw [arrow] (type) -- (val);
        \draw [arrow] (type) |- (eval);
        \draw [arrow] (type) |- (emp);

        \draw [arrow] (val) -| (5.5,0) |- (syn);
        \draw [arrow] (val) -| (5.5,0) |- (sim);

        \draw [arrow] (eval) -| (5.5,-2) |- (off);
        \draw [arrow] (eval) -| (5.5,-2) |- (real);
        
    \end{tikzpicture}
    \caption{Taxonomy of Research Types}
    \Description{Diagram showing the taxonomy of research types including validation research, evaluation research and empirical study. Validation and evaluation types are linked to a second layer classifying the experimental environments. The three environments are simulated closed-loop, real-world environment and offline dataset-based.}
    \label{fig:Research-Types}
\end{figure}
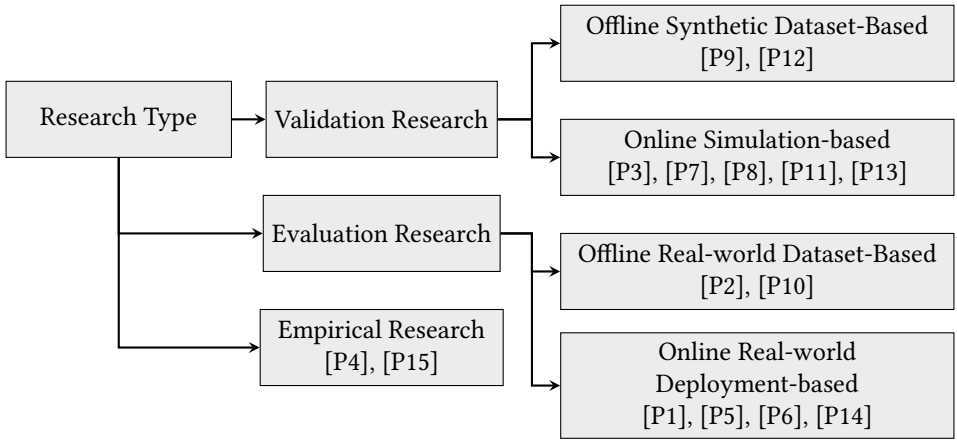

\figurename~\ref{fig:Research-Types} shows the derived taxonomy of research types and includes the associated paper IDs. A detailed explanation of the mapping results is provided in Section~\ref{sec:mapping results}. 

Validation studies utilise lab based methods such as simulation and synthetic databases to develop an idea, whilst evaluation studies utilise real world data and physical environments to show feasibility of methods in deployment. Some experiments only require a portion of the ADS stack to be run in an offline setting. Both validation and evaluation studies can take this approach, with the difference being the source of the dataset. If experiments are run on a dataset of real-world information/recordings, then the study can be considered as an evaluation study; however, if the data is synthetically produced in the lab from simulation or other means, it is a validation study. This categorisation highlights the split between lab based "proof of concept" studies vs those intended to demonstrate how a method or tool can work in a real-world setting.

\subsection{Mapping Results}
\label{sec:mapping results}
Figure~\ref{fig:Research-Types} summarises the classification results using the taxonomy of research types along with their experimental environments.

Evaluation studies \pref{P1, P2, P5, P6, P10, P14} show ADS practicality and reliability in real world settings, demonstrating how a system will interact with its environment when deployed especially in papers which are real world deployment based. On the other hand, validation studies \pref{P3, P7, P8, P9, P11, P12, P13} tend to utilise simulations or lab experiments to present new methods for ADS debugging with room for future industry applications. Both \pref{P4} and \pref{P15} are empirical studies that focus on understanding ADS bugs rather than introducing or evaluating debugging methodologies. \pref{P4} performs a large-scale analysis of historical development artefacts, such as commits, issues, and pull requests to identify and classify ADS bugs. Similarly, \pref{P15} conducts an empirical analysis of bug-fix patterns by extracting commits, issues, pull requests from existing repository to identify real bug-fix patterns in ADS. Therefore, these two papers are empirical studies (see Section \ref{sec:mapping results RQ3}), and do not propose a new debugging technique or evaluate any debugging tool or algorithm. They are therefore not included in answering RQ2: Research Problems (Section \ref{sec:problems}).

Experimental environments across papers appears balanced with the outlier of synthetic dataset based, potentially due to increased work required to produce a specialised dataset. The frequent use of simulation indicates that researchers are investigating ADS debugging in controlled environments that are inspired by real-world scenarios. As a cost-effective approach to test ADS systems, simulation offers a safe and flexible platform, making it especially valuable for prototyping. Studies utilising real world data are common, offering contextual insights that provide a more realistic perspective on how an ADS operates. Experiments on real data demonstrate how particular components of ADS function removing the "black box" appearance to provide further insight. Research in this area involves applying a specific methodology to a particular set of data such as testing a classifier for an ADS \cite{paper13} or utilising an established method on a part of a system \cite{paper4, paper7}. 

We can note from our results the lack of other commonly recognised software engineering study types within ADS debugging. We found no philosophical papers offering new conceptual insights, and no solution-proposal papers without accompanying implementations. This suggests that work in this field tends to include an implemented technique rather than theoretical or conceptual contributions \cite{paper16}. This is not to say the existence of theoretical papers in the field is impossible, but it likely suggests tools for testing ADS are widely available, so constructing an implementation to reinforce the proposed method is feasible in most cases. Additionally, there are no opinion pieces or experience reports, limiting the number of alternative perspectives focused on implementation. A lack of secondary papers accompanies the general lack of studies, suggesting the field's infancy and need for further research.

\begin{figure}
    \centering
\includegraphics[width=0.7\linewidth]{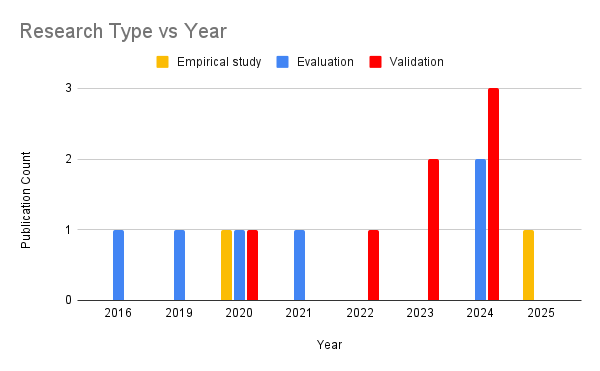}
    \caption{Evolution of various Research Types}
    \Description{Bar chart of evaluation, validation and empirical study types published between the years 2016 to 2024. Evaluation studies are consistently 1 per year apart from 2020 with 2 and 2023 with none. Validation studies are all included in the years 2023 and 2024 with 4 in 2023 and 3 in 2024. There is a single empirical study that was published in 2020.}
    \label{fig:research type}
\end{figure}

Figure \ref{fig:research type} shows that in terms of publication years the number of evaluation studies appear to be consistently low over time demonstrating the need to ensure safety of ADS in development but also showing a potential lack of interest in the field from industry. Limited collaboration with industry may arise from challenges associated with deploying complex ADS in real-world settings, along with budget constraints. Conversely, a significant portion of recent research consists of validation studies, indicating a growing interest in the field among academics. Research in academia often focuses on prototyping new methods prior to any significant investment from industry, implying that we can anticipate a subsequent rise in evaluation studies over the coming years to test newly proposed methods. Our findings revealed few empirical studies \pref{P4, P15}, highlighting the current lack of efforts aimed at identifying trends in this field. It is typical for a developing field to have fewer empirical studies, although a greater prevalence of secondary research papers, such as \pref{P4}, are expected to emerge in the future.

%% file: Sections/Problems.tex
\section{RQ2: ADS Debugging Problems}\label{sec:problems}

\subsection{Taxonomy of Research Problems}

We carefully reviewed all the final papers and extracted new definitions for each of the three main debugging problems from conventional systems (see Section \ref{conventional-debug}). ADS debugging problems are similar to conventional debugging problems presented, but are specifically tailored for the complexities of ADS. 
We also found localisation was not as simple to define from our results as some papers aim for fault triggering conditions in the environment, whilst others search for ADS code-level faults. We accounted for this by proposing \textit{simplification} as another definition to describe problems that appear as localisation but are caused by environmental triggers rather than code faults. The definitions of the ADS debugging problems are as follows.

\begin{definition}[{ADS Scenario Simplification}]
For an autonomous driving system $S$ and its failure or anomalous behaviour $f$ observed under a driving scenario defined by a set of scenario entities $E = \{e_1, \ldots, e_n\}$, the problem of \textit{ADS scenario simplification} is to find a minimal subset $E' \subseteq E$ such that $f$ still occurs under $E'$. This includes:

(i) Entity-level simplification, which focuses on \textit{environmental entities} (e.g. vehicles, pedestrians, obstacles, road agents) that are responsible for inducing failure, and removes entities that are not relevant to the test failure. This reduces processing load, i.e. repeat tests for localisation will take less time and also highlights remaining potential fault triggers, providing insights for the downstream debugging tasks. 

(ii) Frame-level simplification, which focuses on \textit{environmental entities} whose corresponding temporal \textit{frames} triggered failure. This reduces the effective scenario length, allowing the developer to focus on the fault-relevant frames without reviewing lengthy recordings.
\end{definition}

\begin{definition}[ADS Fault Localisation]
For an autonomous driving system $S$, consisting of structural elements $C = \{c_1, \ldots, c_n\}$ (e.g. modules, files, or lines of source code), and an observed failure or anomalous behaviour $f$, the problem of \textit{ADS failure-inducing elements localisation} is to find a minimal subset $C' \subseteq C$ such that $f$ still occurs only if the elements in $C'$ are retained. This includes: 

(i) Module-level localisation isolates faulty high-level modules such as perception, planning, or control that are responsible for the failure.

(ii) Message-level localisation isolates the specific output messages (e.g. sensor messages, planner commands) of module executions that induce the failure. 

(iii) Code-level localisation pinpoints the specific regions of source code (such as statements, parameters, or integration rules) that induce failure.

\end{definition}

\begin{definition}[ADS Failure Explanation]
For an autonomous driving system $S$ and an observed failure or anomalous behaviour $f$, the problem of \textit{ADS failure explanation} is to understand why $f$ occurs and how it propagates within the system by producing a human-understandable narrative.
\end{definition}

\begin{definition}[ADS Failure Repair]
For an autonomous driving system $S$ with faulty elements $C$ that trigger an observed failure or anomalous behaviour $f$, the problem of \textit{ADS automatic failure repair} is to repair $C$ with a minimal set of modifications $M$ such that applying $M$ to $C$ produces a modified system $S'$ where $f$ is no longer exhibited. 
\end{definition}

\begin{figure}
    \centering
    \scalebox{0.8}{
    \begin{tikzpicture}[node distance=1.5cm]
        \node (start) [process] {ADS Debugging Problem};
        
        \node (sim) [process, right of=start, xshift=2.5cm] {Simplification};
        \node (ent) [process, right of=sim, xshift=4cm] {\parbox{5cm}{\centering (Relevant) Environmental Entities \\ \pref{P7}, \pref{P11}}};
        \node (fra) [process, below of=ent] {\parbox{3cm}{\centering (Relevant) Frames \\ \pref{P1}, \pref{P13}}};

        \node (loc) [process, right of=start, xshift=2.5cm, yshift=-3cm] {Localisation};
        \node (mod) [process, right of=loc, xshift=3cm] {\parbox{3cm}{\centering Module \\ \pref{P5}, \pref{P8}, \pref{P10}, \pref{P14}}};
        \node (mes) [process, below of=mod] {\parbox{3cm}{\centering Message \\ \pref{P10}, \pref{P11}, \pref{P13}}};
        \node (cod) [process, below of=mes] {Code};
            \node (par) [smallprocess, right of=cod, xshift=2.5cm] {\parbox{3cm}{\centering Parameters \\ \pref{P6}}};
            \node (sta) [smallprocess, above of=par, yshift=-0.3cm] {\parbox{3cm}{\centering Statements \\ \pref{P1}, \pref{P12}}};
            \node (int) [smallprocess, below of=par, yshift=0.3cm] {\parbox{3cm}{\centering Integration Rules \\ \pref{P3}}};

        \node (exp) [process, right of=start, xshift=2.5cm, yshift=-9cm] {Explanation};
        \node (nar) [process, right of=exp, xshift=4cm] {\parbox{3cm}{\centering Temporal Narrative \\ \pref{P2}, \pref{P9}, \pref{P13}}};

        \node (rep) [process, right of=start, xshift=2.5cm, yshift=-12cm] {Repair};
        \node (cod2) [process, right of=rep, xshift=3cm] {Code};
            \node (par2) [smallprocess, right of=cod2, xshift=2.5cm] {\parbox{3cm}{\centering Parameters \\ \pref{P6}}};
            \node (sta2) [smallprocess, above of=par2, yshift=-0.3cm] {\parbox{3cm}{\centering Function/Method \\ \pref{P12}}};
            \node (int2) [smallprocess, below of=par2, yshift=0.3cm] {\parbox{3cm}{\centering Integration Rules \\ \pref{P3}}};

        \draw [arrow] (start) -- (sim);
        \draw [arrow] (start) |- (loc);
        \draw [arrow] (start) |- (exp);
        \draw [arrow] (start) |- (rep);

        \draw [arrow] (sim) -- (ent);
        \draw [arrow] (sim) |- (fra);

        \draw [arrow] (loc) -- (mod);
        \draw [arrow] (loc) |- (mes);
        \draw [arrow] (loc) |- (cod);
            \draw [arrow] (cod) -- (par);
            \draw [arrow] (cod) -- (sta);
            \draw [arrow] (cod) -- (int);

        \draw [arrow] (exp) -- (nar);

        \draw [arrow] (rep) -- (cod2);
            \draw [arrow] (cod2) -- (par2);
            \draw [arrow] (cod2) -- (sta2);
            \draw [arrow] (cod2) -- (int2);
    \end{tikzpicture}}
    \caption{Taxonomy of Debugging Problems. Some papers are listed multiple times when they address more than one problem. \protect\pref{P4} and \protect\pref{P15} were excluded from the analysis because they constitute empirical research.}
    \label{fig:problem-taxonomy}
\end{figure}
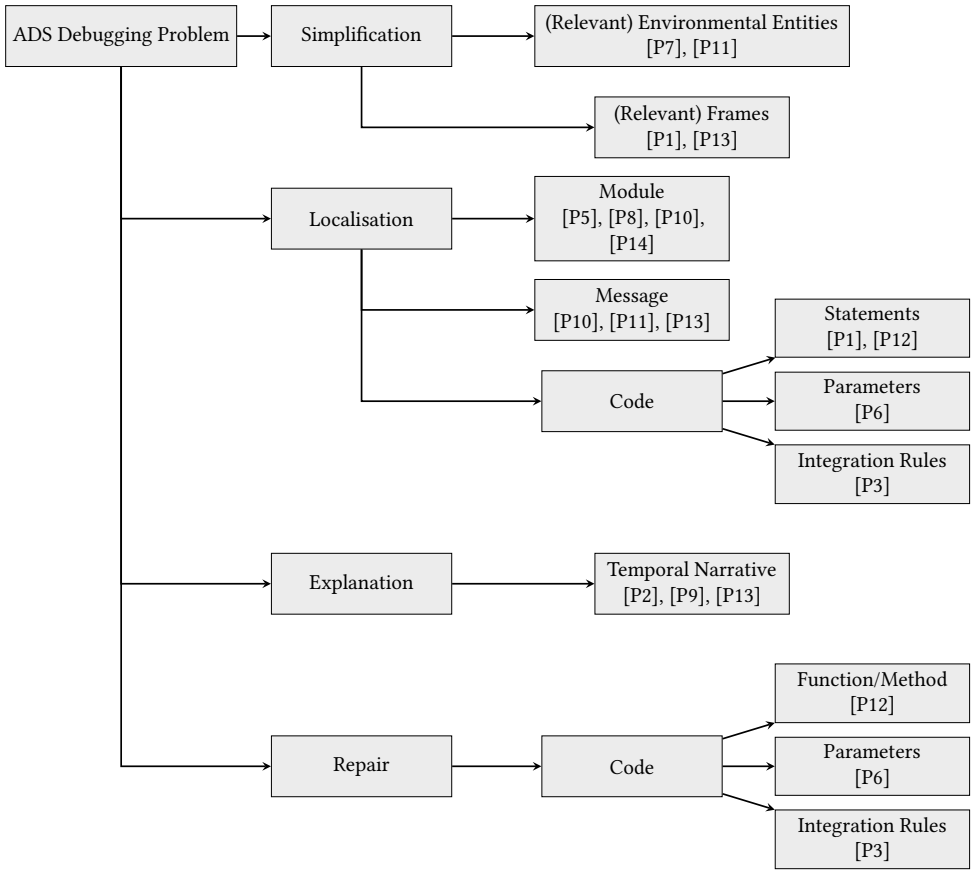

In addition to each of the proposed ADS problem definitions, we found that each problem was addressed in a number of ways as shown in Figure \ref{fig:problem-taxonomy} and the mapping results explained in Section \ref{sec:mapping}. For example, the simplification problem targets either the environmental entities (e.g. surrounding vehicles) or the recording frames of the scenario. By analysing which aspects each problem targets, we can identify potential starting points for future research and highlight areas of ADS debugging that have been explored versus those that remain under-investigated.

\subsection{Mapping Results}
\label{sec:mapping}
By proposing the definitions for each of the debugging problems specific to ADS, we can individually discuss the work on each problem and provide clear definitions for future ADS debugging studies. Our mapping results show how each debugging task can be applied to different areas of ADS with a full list of associated papers for each problem as shown in Figure \ref{fig:problem-taxonomy}.

\subsubsection{Simplification}
The Simplification problem acts as more of a precursor task to the other debugging problems, with explicit focus on reducing the problem space for the following localisation, explanation, and repair tasks. Generated test scenarios contain many vehicles and environmental entities that do not contribute to the observed failure. This makes debugging difficult, as it becomes unclear which agents are actually influencing the behaviour of the autonomous vehicle. The problem of scenario simplification is to reduce a test scenario to a minimal set of scenario entities responsible for triggering a failure using entity-level or frame-level simplification approaches. 

Two studies \pref{P7, P11} have explored the problem of reducing a complex ADS scenario to the minimal elements to reproduce the problem. \pref{P7} addresses the problem of scenario simplification at the entity level by reducing the traffic participants to only those required for the failure to occur. By eliminating irrelevant participants while preserving the failure, this reduces developer workload, accelerates repeated test executions, and makes the fault-triggering interactions more explicit. Similarly, ROCAS \pref{P11} isolates only the accident-triggering entities and discards those that have no causal influence on the failure, effectively reducing to minimal set of environmental entities required for the failure to occur before performing further localisation steps. 

The studies \pref{P1, P13} take an alternative approach to scenario simplification by focussing on the simulation and frame data rather than participants in the scenario. \pref{P1} addresses scenario simplification at the frame level by serializing the exact internal state of software components at the moment the failure occurs. This allows developers to reload and replay only the critical execution cycle(s) surrounding the fault, effectively discarding all preceding frames that do not contribute to the failure. In doing so, P1 reduces the effective scenario length and focuses debugging effort on the temporal frames that matter. Likewise, the ACAV framework described in \pref{P13} automatically segments long accident recordings and discards frames that are irrelevant to the observed failure, based on map features, detected NPCs, prediction outputs, and planning decisions. By reducing the multi-second recordings to only the safety-critical temporal windows in which the violation develops, allowing engineers to focus exclusively on frames that contributed to the accident.

\subsubsection{Localisation}

Localisation problems aim to identify the root cause of a fault. Some approaches seek to identify faults at the module level, such as \pref{P14}, which identifies faulty modules to direct manual debugging efforts. Similarly, \pref{P5} analyses the behaviour of the path-planning module under different parameter settings to detect faulty planner configurations, and \pref{P8} explains incorrect lane-change decisions by attributing them to misbehaving perception or learning-based decision modules. 

Other observed papers in the literature focus on message-level localisation, where the goal is to pinpoint which specific inter-module message or computation result induces the violation. For example, \pref{P13} analyses the messages exchanged between the perception, prediction, and planning modules, discards irrelevant frames, and identifies safety-critical messages that lead to violations. In contrast, \pref{P11} uses counterfactual re-execution to determine which module output first deviates from the reference execution, isolating both the triggering entity and the module producing the faulty internal state. \pref{P10} identifies the ADS module that caused the violation, and then isolates the specific message within that module that triggered the faulty behaviour. Thus, it performs both module-level and message-level fault localisation.

At a finer granularity, code-level localisation aims to identify the exact code statements, parameters, or integration rules that cause failures. For example, \pref{P6} identifies fault-inducing system parameters within the ADS path-planning cost model and \pref{P3} identifies faulty integration rules whose incorrect ordering or conditions lead to feature-interaction failures, directly revealing the code regions requiring repair. \pref{P1} helps developers identify the specific source-code statements that cause the failure by serialising the exact internal state of the ADS module at the moment of failure and replaying it in a step-through debugging environment. This allows the faulty execution path, variable values, and control-flow transitions to be examined exactly as they occurred, ultimately isolating the precise line of code responsible for the defect. \pref{P12} narrows localisation down to the specific functions and implementation details within the ADS control algorithms by using assertion-guided analysis to trace anomalous behaviours back to the internal functions that generate them.

\subsubsection{Explanation}
Unlike localisation, approaches to explanation problems don't seek to find what caused a failure, but how and why. We view examples of this in the literature with temporal narratives which aim to describe an ADS failure in a time-stepped manner. \pref{P2} explains why an autonomous race car crashed by analysing the sequence of events leading up to the failure, examining perception outputs, planner decisions, vehicle dynamics, and environmental factors. \pref{P13} describes the events in order of occurrence. Causality can be attributed to parts of the ADS, which may have been incorrect at that time frame. Explanatory insights are a key step in any debugging effort, as they increase developers' understanding of a system's decision process. Explanation methods are also relevant to systems that rely on machine learning, with more studies aiming to explain the confusing nature of ML misclassification. For instance, \pref{P9} investigates a lane changing model by generating explanations for its decisions leading to failures.

\subsubsection{Repair}
The repair problem builds on the localisation problem by aiming to correct identified faults. This is generally done at the code level, following a similar taxonomy structure to the localisation model, focusing on either parameters or integration rules. Parameter repair involves changing the assigned parameters of a system to address incorrect behaviour without introducing new faults, turning the parameter-based approach into an optimisation problem \pref{P6}. Integration rule repair follows on from the localisation of incorrect logic in rule order. Repair methods seek to rearrange the integration rules via mutation to find combinations that do not exhibit problematic behaviour \pref{P3}. An addition to the code level taxonomy diagram for repair is function/method repair. We observed method repair in \pref{P12}, which presents a method for predicting code logic during development, allowing developers to catch logic failures before fully implementing them. DNN-related methods also apply to repair problems, as many papers have observed, by automatically repairing classifiers for use in ADS. Most of these methods apply more generally to machine learning, but the importance to ADS should be noted.

%% file: Sections/Techniques.tex
\section{RQ3 Results: Techniques}\label{sec:techniques}

\subsection{Taxonomy of Techniques}

For RQ3, we compiled a list of all distinct debugging techniques proposed, including fault injection and causal analysis. We tracked the frequency and applications to understand which methods are common across problem types. A taxonomy of problem categories was devised to represent overlaps in the literature and highlight similar approaches.

\begin{figure}
    \centering
    \scalebox{0.8}{
    \begin{tikzpicture}[node distance=1.2cm]  
        \node (start) [process] {Techniques};

        \node (one) [process, right of=start, yshift=2.4cm, xshift=6cm] {\parbox{5cm}{\centering Manual/Empirical Analysis \\ \pref{P2}, \pref{P4}, \pref{P15}}};
        \node (two) [process, below of=one] {\parbox{5cm}{\centering Causal/Assertion Analysis \\ \pref{P9}, \pref{P10}, \pref{P12}, \pref{P13}, \pref{P14}}};
        \node (three) [process, below of=two] {\parbox{5cm}{\centering Fault Injection/Mutation \\ \pref{P3}, \pref{P8}, \pref{P11}}};
        \node (four) [process, below of=three] {\parbox{5cm}{\centering Iterative Minimisation \\ \pref{P1}, \pref{P6}, \pref{P7}}};
        \node (five) [process, dashed, below of=four] {\parbox{5cm}{\centering ML-Debugging (\S \ref{sec:discussion:ml-debugging}) \\ \pref{P5}}};

        \draw [arrow] (start.east) -- (one.west);
        \draw [arrow] (start.east) -- (two.west);
        \draw [arrow] (start) -- (three);
        \draw [arrow] (start.east) -- (four.west);
        \draw [arrow] (start.east) -- (five.west);
    \end{tikzpicture}}
    \caption{Taxonomy of Technique Categories. The dashed box indicates ML-debugging techniques that are not specialised for ADS debugging.}
    \Description{A single layered taxonomy diagram displaying the 5 technique categories of Manual/Empirical Analysis, Causal/Assertion Analysis, Fault Injection/Mutation, ML-Debugging and Iterative Minimisation.}
    \label{fig:Techniques}
\end{figure}
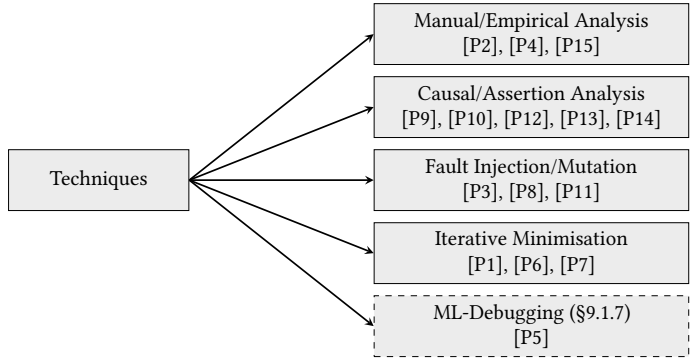

Our taxonomy of all observed technique categories is shown in Figure \ref{fig:Techniques}, and the mapping results are explained in Section \ref{sec:mapping results RQ3}. Techniques from each paper were generalised to allow grouping and identify trends. Each approach is based on one of five criteria, whilst the specific techniques used in each paper are listed in our data extraction results \cite{SMSdataset}. Despite being out of scope for the general taxonomy, ML-Debugging methods can be applied to ADS and are worth noting as an additional technique criterion (see \S \ref{sec:discussion:ml-debugging} for further discussion). Each technique can be defined as:

\begin{itemize}[-]
    \item \textit{Manual/Empirical Analysis}: Direct analysis of a system by a developer.
    \item \textit{Causal/Assertion Analysis}: Checking causal links within the system, potentially using assertions.
    \item \textit{Fault Injection/Mutation}: Manipulation of the faults and code to gain insights and generate patches.
    \item \textit{Iterative Minimisation}: Repeating tests with alternative setups to reduce scenarios to fault triggering conditions.
    \item \textit{ML-Debugging}: Techniques to repair a neural network to correct predictions. 
\end{itemize}

\subsection{Mapping Results}
\label{sec:mapping results RQ3}
Figure \ref{fig:Techniques} presents the papers grouped by techniques from our previously defined taxonomy.

\paragraph{Manual/Empirical Analysis}
Some papers rely more on conventional software debugging methods, applying a manual approach to investigating code faults. Manual debugging is applicable to complex systems like ADS, as evidenced by the works of \pref{P2}, which implement manual investigation using the logs of an autonomous vehicle post-incident. This manual investigation of ADS data provides insights for developers into how a failure can be logged and deciphered, providing a guideline for investigating future failures. \pref{P4} completes a similar manual analysis of ADS faults within their work, providing details on bugs and their symptoms from multiple projects as opposed to a particular method for an individual system. \pref{P15} similarly performs a large-scale empirical analysis, focusing on characterising bug-fix patterns in ADS through manual inspection of bug-fix pull requests. These studies provide insights regarding how developers repair ADS faults and common strategies used in practice. Such techniques can potentially draw inspiration from traditional techniques explored in Section \ref{conventional-debug}.

\paragraph{Causal/Assertion Analysis}
Following an ADS failure, it is essential to track the cause of the problem to reveal and understand it. Unlike manual code analysis, some approaches seek to use more automated methods to trace the causality of a failure. For example, \pref{P10} implemented a causal analysis technique, using counterfactual reasoning within simulation to establish causal links between module outputs and system-level driving violations. This study helps in identifying the ADS module responsible for the violation, and pinpoints the specific message that induces the violation.
\pref{P13} presents ACAV, an automated framework that identifies safety-critical events in two stages. The first stage removes frames of the scenario by removing irrelevant frames before and after the event, reducing the search space. Following this a causal analysis tool (CAT) based on safety specifications is applied to identify specific events contributing to the failure of the ADS.
Another method, presented in the work of \pref{P14}, utilises a modular fault diagnosis framework tailored to the structure of ADS, enabling fault isolation by analysing failures across system components.
Many of the papers present new techniques to automate the analysis process, suggesting the complexity of ADS calls for tools to improve efficiency. Another type of automated method observed is the generation of explanations linked to root causes. As shown in \pref{P9}, a partially automated analysis method can assist in understanding failures rather than just finding them. Tools which actively generate symptom explanations for a developer will improve the understanding of the problem, thus decreasing the time needed to generate a repair. Reducing debugging time can also be achieved by locating logic errors early as shown by \pref{P12}, which uses assertion rule generation to predict model behaviour in early development, highlighting errors before they are implemented. This paper highlights a need for methods to analyse the causes and impacts of failures to make the process more reliable, as well as contribute to a deeper understanding of ADS debugging.

\paragraph{Fault Injection/Mutation}
A recurring technique in ADS debugging involves using mutation or fault injection to modify the system and expose failures or weaknesses. By modifying failure-inducing inputs, developers can observe how the system responds, which helps identify the root cause of the failure.
An example is provided by \pref{P11}, which applies mutation to both driving scenarios and internal components of the ADS to uncover distinctive behaviours linked to specific root causes.
Another example of mutation can be seen in \pref{P3}, in which the interactions between pre-coded control algorithms are mutated to avoid failing scenarios. 
Monitoring the system's reaction to a known failing input also provides valuable insights for future debugging, particularly when similar failures arise during live testing. This approach allows developers to trace system behaviour as faults are injected, making it easier to recognise similar issues in real-world operation. 
Fault injection into an ADS is used in the works of \pref{P8}, with faulty data being injected into the camera system. Other perception sources are used against this data to check and identify faults in sensor data and filter accordingly.
Fault manipulation methods are commonly introduced as testing methodologies; however, incorporating additional steps to further investigate each generated failure can significantly enhance the debugging process.

\paragraph{Iterative Minimisation}
Iterative techniques are present within ADS debugging, repeating a combination of modification and testing of an ADS environment until a root cause fault is converged upon. One example is seen in the work of \pref{P7}, which seeks to reduce scenario input to limit simulations to only key components. A further debugging step is explored to show how fault triggers can be identified by finding the components of the environment which, when removed, cause the failure to no longer be observed. Identifying fault triggers could be used to locate internal faults by providing the developer with specific failure-inducing input. Iterative techniques can also be applied internally to ADS as seen in the method proposed by \pref{P1}. Serialising, or saving the states of ADS at a failure, allows the failure to be recreated in simulation. By repetitively running the failing scenario with modified internal states, a perfect state can be found in which the failure is no longer observed. When a correct internal state is found, the original ADS can be modified to avoid previously encountered failure-inducing states.
Iterating over parameters can also serve a debugging purpose, with the approach presented in \pref{P6} iterating over input to key decision parameters to expose faults. Once faults are exposed the parameters linked to the fault are used to provide key insights into why the system failed and suggest what needs to be adjusted to avoid future failing scenarios.

\paragraph{ML-Debugging}
Techniques specific to machine learning can be applied to ADS that implement ML models. Many systems will utilise a classifier at the perception stage to translate information on the environment into a form that the path planner can use. Some debugging papers, such as the works of \citet{paper11,paper13}, are particularly relevant to ADS, since they show how deep neural networks (DNN) patching can be applied to classifiers used for perception. Re-training and patching are both methods which can be utilised to repair ML models which are part of a system. Our classification for ML techniques doesn't just include methods to repair classifiers incorporated into ADS, but also the use of machine learning for otherwise standard debugging tasks. The approach in \pref{P5} makes use of a neural network as part of a diagnostic pipeline to detect and diagnose faults in operation. The use of machine learning models and techniques for standard debugging tasks highlights an opportunity to utilise ML in more ADS debugging tasks.

It is also important to note that the methods vary depending on the development stage of the ADS. Some approaches are designed for \textit{in-development debugging}, aiming to detect faults during production of the system \pref{P1, P3, P5, P6, P7, P12, P13}, while others are intended for \textit{post-deployment debugging}, targeting failures observed in deployed systems \pref{P2, P8, P9, P11, P14}. Post-deployment debugging often requires more generalizable methods, as the developers performing the analysis may not have been involved in the original system design. Understanding this distinction helps in selecting appropriate debugging strategies for different stages of ADS development and deployment.

\subsection{Other Problem-Specific Techniques}
Aside from the previously discussed classifications, a wide range of problem-specific techniques were observed in the literature. Techniques often align with one of the proposed problem categories (simplify, localise, explain or repair). These techniques target a specific level, aligning with the previously defined taxonomy.

\paragraph{Simplification Techniques}
These papers apply techniques (typically within simulation) to reduce the search space of a scenario whilst preserving the fault trigger. In their work, \pref{P7} introduce scenario simplification, which is similar to delta debugging, where aspects of the scenario are removed incrementally \cite{delta-debug}. If the failure is still observed after the removal of a non-ego vehicle, then it is permanently removed, reducing the scenario to only cars which trigger the incorrect behaviour. Similarly, the study by \pref{P13} proposes causality analysis to identify causal relationships between the failure and events leading to it. A simplification technique is used to reduce scenario recording to only contain aspects of the environment with causal links to the failure by reducing frames before and after causal events appear. Together, these works demonstrate how debugging can also apply in a scenario setting by searching for fault triggers to highlight faults, rather than direct fault localisation. These simplification methods play a significant role in not only reducing the search space but also highlighting the fault triggers which can be used to direct a localisation technique.

\paragraph{Localisation Techniques}
Techniques designed for localisation problems vary in granularity, with some focused on highlighting sections of code containing faults while others point to specific lines of code. In their work, \pref{P1} introduces logging and tracing to reproduce failures, allowing developers to analyse execution paths during development. Other approaches are more direct, such as the implementation by \pref{P11}, which applies mutations to the simulated environment and ADS internal states simultaneously. This cyber-physical co-mutation helps replicate failures in a controlled manner to trace faults that may have contributed to the observed issue. Highlighting the specific misconfiguration in ADS remains a common goal among applied techniques. A notable example is the work of \pref{P12}, which presents a unique method to achieve this goal by modelling ADS behaviour during development.

\paragraph{Explanation Techniques}
Explanation techniques were primarily applied to the system as a whole and are useful for explaining scenario failures or interpreting machine learning models. One technique proposed by \pref{P9} uses entropy-based value selection to assist their proposed tool, SHAP (SHapley Additive exPlanations), in interpreting the decisions of a lane changing model. When multiple valid Shapley explanations are generated, the one with the highest entropy is selected, which is the explanation with the most uniformly distributed responsibility across features. Failure explanations generated enable engineers to easily understand the scope of a failure and identify key features influencing failing scenarios. In another instance, a structured analysis of crash logs generated by an ego vehicle is presented in the works of \citet{paper1}. While not providing an automated technique to assist debugging, the proposed guidelines provide a standard framework for future manual analysis of ADS failures. Future failure analysis can take inspiration from this work, utilising guidance on creating crash logs for traceability and observational reasoning. More structured notation can help to assist in developing the interpretability of failures and set future standards. 

\paragraph{Repair Mechanisms}
Repair techniques in ADS include methods of patching faults or reconfiguring system behaviour based on identified issues. The work of \pref{P3} address the automated repair of feature interaction failures. When referring to feature interactions in ADS, we refer to the interaction of various algorithms responsible for tasks such as lane-keeping or pedestrian avoidance. These behaviour algorithms are located within the control module, which prioritises them in a predefined order. In this context, \pref{P3} propose the use of a genetic algorithm with mutation-based search to explore alternate system priorities to avoid undesired behaviour as a result of suboptimally ordered control algorithms. DNN repair is a notable part of the repair problem with application to ADS, drawing on techniques from the much larger field of machine learning. Two papers by \citet{paper13} and \citet{paper11} focus on the repair of DNNs and utilise similar techniques. In their study, \citet{paper13} use DNN fault localisation and Arachne (search-based patching) to repair faulty weights in a model. A search-based optimisation method is also utilised to find a weight correction with the best possible outcome. Similarly, \citet{paper11} use a set of techniques but with a modified patching technique (Adaptive Arachne). Both these papers take methods from general ML debugging and apply them to an ADS.

%% file: Sections/Tools.tex
\section{RQ4: Tools} \label{sec:tools}

While ADS grow increasingly complex, diagnosing and resolving faults becomes increasingly challenging. To simplify the landscape of currently available support in ADS debugging, we extracted the tools used within each study in Table \ref{tab:tools-table}, listing both existing and custom-made tools and their use cases. We note each tool's name along with its input, output, features and link to the tool. Tools vary greatly in scope from fault localisation tools to ADS simulators for scenario testing. This summary provides a practical reference for those seeking tools to assist ADS debugging by presenting available software in a simplified form. 

\begin{table}
    \centering
    \caption{List of supporting Tools and Simulation Environments for ADS Debugging}
    \scriptsize{
    \begin{tabularx}{\linewidth}{p{1.5cm}p{2cm}p{2cm}Xp{2cm}p{2cm}}
        \toprule
        \textbf{Tool} & \textbf{Inputs} & \textbf{Outputs} & \textbf{Features} & \textbf{Link} & \textbf{Category} \\
        \midrule
        SHAP & ML model & Feature attributions & Model interpretation tool based on Shapley values & \url{https://github.com/shap/shap} & Tool \\
        Matlab/Simulink & Simulation models, crash data & Diagnostic results	& Widely used numerical computing platform for simulation and analysis & \url{https://www.mathworks.com/products/matlab.html} & Tool \\
        ACAV & Accident recordings & Causality analysis results & Fault diagnosis framework that automates causality analysis in accident data & \url{https://github.com/username/ACAV} & Tool \\
        XGBoost C++ Library & System logs, serialized system states & Fault predictions, reproducible simulation states & Combines ML-based fault detection with state serialization for replay/debugging & \url{https://github.com/dmlc/xgboost} & Tool \\
        Adaptive Search-based Repair (AdRep) & Trained Deep Neural Network, Faulty model weights, Training data & Repaired Deep Neural Network with improved accuracy & Adaptive search-based fault localization and repair of DNN weights & \url{https://github.com/jst-qaml/adaptiveRepair} & Tool \\
        Apollo & ADS model, Scenario & Simulation of driving scenario & Simulation platform for testing ADS & \url{https://github.com/ApolloAuto/apollo} & Simulation Environment \\
        LGSVL & ADS model, Scenario & Simulation of driving scenario & High fidelity simulator for ADS & \url{https://github.com/lgsvl/simulator} & Simulation Environment \\
        Autodrive & ADS model, Scenario & Simulation of driving scenario & Simulation platform & \url{https://github.com/AutoDRIVE-Ecosystem} & Simulation Environment \\
        PreScan & ADS model, Scenario & Simulation of driving scenario, focusing on sensor simulation & Commercial ADS simulator & \url{https://plm.sw.siemens.com/en-US/simcenter/autonomous-vehicle-solutions/prescan/} & Simulation Environment \\
        CARLA & ADS model, Scenario & Simulation of driving scenario & Open-source ADS simulator &\url{https://github.com/carla-simulator/carla} & Simulation Environment \\
        \bottomrule
    \end{tabularx}
    }
    \label{tab:tools-table}
\end{table}

Most readily available tools are simulators for running ADS scenarios as noted in Table \ref{tab:tools-table}, reflecting the importance of scenario-based evaluation. As well as simulators, tools for causality analysis, neural network repair, model interpretability and more are observed. Each tool targets a specific debugging task with little competition from alternative approaches to the same problem. A general lack of available implementations comes as a result of the field's immaturity, with most papers presenting frameworks for debugging without sharing the coded implementation. As many papers present a framework or guidelines for debugging ADS, we felt it necessary to compile a list of papers with their specific guidance. Table \ref{tab:frameworks-table} presents each paper that can work as a guide for a specific ADS debugging task. For each guide, we present a simplified purpose to help readers identify resources for specific debugging tasks.

\begin{table}
    \centering
    \caption{List of ADS Debugging Tools and Guides}
    \scriptsize{
    \begin{tabularx}{\linewidth}{p{4cm}Xp{4cm}}
        \toprule
        \textbf{Framework/Guide} & \textbf{Paper Title} & \textbf{Purpose} \\
        \midrule
        Bug classification framework & A Comprehensive Study of Autonomous Vehicle Bugs \cite{paper5} & Categorizes and analyses AV software bugs \\
        Fault diagnosis guidelines & Autonomous Driving—A Crash Explained in Detail \cite{paper1} & Suggests process for analysing AV system failure \\
        Space-Time Fusion Framework & Fault Diagnosis of the Autonomous Driving Perception System Based on Information Fusion \cite{paper18} & Fuses spatial and temporal data for fault identification \\
        Scenario Simplification Framework & Less is More: Simplification of Test Scenarios for Autonomous Driving System Testing \cite{paper16} & Reduces scenario complexity to isolate faults \\
        ACAV / CAT & ACAV: A Framework for Automatic Causality Analysis in Autonomous Vehicle Accident Recordings \cite{paper15} & Determines causal links between components and failures \\
        ROCAS & ROCAS: Root Cause Analysis of Autonomous Driving Accidents via Cyber-Physical Co-mutation \cite{paper9} & Identifies fault propagation paths via co-mutation \\
        ADAssure & ADAssure: Debugging Methodology for Autonomous Driving Control Algorithms \cite{paper14} & Offers structured debugging of control algorithms \\
        Module Diagnostic Framework & Modular Fault Diagnosis Framework for Complex Autonomous Driving Systems \cite{paper3} & Modular diagnosis of fault locations in ADS pipeline \\
        Parameter Coverage Criteria & Parameter-Based Testing and Debugging of Autonomous Driving Systems \cite{paper6} & Guides testing/debugging based on parameter combinations \\
        ARIEL & Automated Repair of Feature Interaction Failures in Automated Driving Systems \cite{paper4} & Framework for detecting and repairing feature interaction faults \\
        DVCA & Towards Automated Driving Violation Cause Analysis in Scenario-Based Testing for Autonomous Driving Systems \cite{paper20} & Framework for code level causal analysis for ADS in simulation \\
        \bottomrule
    \end{tabularx}
    }
    \label{tab:frameworks-table}
\end{table}

These frameworks and guidelines offer valuable perspectives on how to approach various ADS debugging tasks, ranging from fault classification and causal analysis to scenario design and module-level diagnosis. Beyond practical debugging, some guides also enhance understanding of how different types of bugs manifest. By highlighting common patterns these contributions provide a conceptual foundation for future tool development.

%% file: Sections/Discussion.tex
\section{Discussion}\label{sec:discussion}

Our mapping study highlights several overarching trends and challenges in the current state of ADS debugging. This section discusses broader insights that emerged across studies, reflecting the current technical and conceptual maturity of the field. Mapping these trends exposes current limitations of the field and highlights areas for future work.

\subsection{Findings}
\subsubsection{Application Level}
While categorising problem types, it was noted that there is a significant disparity between the application level of approaches to problems. The application level refers to the nodes in the taxonomies detailing how each problem is approached. Scenario simplification typically removes entities or frames, whereas explanation methods describe causal relations within scenarios or employ DNN-based explanation techniques. In contrast, localisation and repair are closely linked, as repair requires localisation to function. Repair methods target the code level, as the code itself must be modified to fix a fault. However, localisation does not follow this trend, with some papers working at the message \pref{P11} and module \pref{P14} levels instead. Higher-level localisation helps developers identify regions of faulty code, but manual effort is still required to pinpoint the lines of code that cause failures. This reflects both search-time constraints and a lack of code-level search-based methods to handle the complexity of ADS.

\subsubsection{Phase}
Our data \cite{SMSdataset} shows strong emphasis on debugging during development, with comparatively few studies addressing post-deployment scenarios. This trend toward development-time debugging more closely aligns with standard software development cycles, indicating proactive efforts to identify and repair software faults before deployment. One possible reason for this trend is the strict guidelines for developing safety-critical software, as ADS systems must pass rigorous testing phases that are likely to reveal numerous failing test cases. Testing will also need to be repeated whenever a fix is applied, which is costly for safety-critical systems. As previously discussed, the lack of debugging relative to testing indicates a research gap: post-deployment debugging. Post-deployment debugging methods differ from development-phase debugging methods because they are less specialised and must approach and modify an already complete system, such as the framework presented in \pref{P13}. While these methods may be less specialised, they are typically more flexible, and several studies have shown that development-phase techniques can be adapted for use after deployment. As a result, both development-time and post-deployment debugging remain promising directions for future research.

\subsubsection{Traditional Software Debugging vs. ADS Debugging}
Debugging autonomous driving systems differs fundamentally from debugging traditional software due to their cyber-physical, environment-dependent, and multi-module architecture. Traditional software typically functions as a standalone system, whereas ADS are required to interact continuously with complex surrounding environments. This ongoing interaction introduces additional behavioural influences beyond the initial input, significantly increasing debugging complexity and necessitating more advanced debugging approaches. Interactions with other software systems, such as simulators and operating systems (e.g., Windows or macOS), can also affect the determinism of ADS. Because ADS testing and debugging rely on input order and timing, and on unintentionally nondeterministic simulators \cite{9793395,11052713}, repeating the same test scenario may not always yield identical outputs, making the deterministic reproduction of failures challenging. Even low-level operating system behaviours, such as thread scheduling or file access tasks, can affect the performance of an ADS \pref{P1}. Any debugging approach must therefore account for sources of nondeterminism that are absent or far less observed in traditional software debugging.

ADS uses a multi-module architecture, with dedicated modules for each driving task, thereby introducing complex dependencies within the system. One issue introduced by module dependencies is error propagation. Small failures in upstream components manifest later in the pipeline, making fault localisation difficult. ADS has observable output at the end of the pipeline in the form of executed actions, and even with recent approaches such as DVCA, it remains difficult to directly link a system-level failure to the output of a particular module at a specific time step \pref{P10}. Future work could investigate techniques for extracting data from ADS pipelines via internal observability, potentially revealing outputs from specific modules during debugging. An approach such as this could, for example, display the exact outputs of the perception module as an annotated environment with identified safety threats, or show the results of localisation by displaying where the ADS believes itself to be in the environment.

Incorrect configurations are observed to account for a substantial proportion of ADS faults \pref{P4}, with failures extending beyond typical software logic errors. Studies show that configuration errors and complex algorithmic faults constitute a large proportion of ADS failures \pref{P4}. These faults often arise from the need to implement complex algorithms, which extend beyond standard software systems. Complex and highly dependent ADS modules also explain why incorrect algorithm implementations account for a notable proportion of ADS failures \pref{P4}. System complexity also increases debugging effort, as searching through lines of code is time-consuming and not only localising the bug but also repairing the issue is complicated. In the study by \pref{P4}, it is noted that implementation errors are rarely trivial to fix, with an average of 104 lines of code modified per repair. Automated repair techniques will need to consider this complexity or rely on manual input.

Empirical findings also reveal that failures frequently concentrate in the planning module, the most safety-critical component of ADS \pref{P4}. Since the planning module determines the ego vehicle’s trajectory, debugging efforts that prioritise this stage can yield substantial safety benefits. Parameter-based localisation techniques such as those proposed in \pref{P6} show that planning-specific debugging can benefit from analysing parameters used in determining speed or trajectory. By defining parameter coverage criteria, their method helps identify misconfiguration or incorrect logic. These findings suggest that future debugging work may benefit from more in-depth targeting of the planning module, potentially integrating parameter-informed guidance into fine-grained code-level localisation.

\subsubsection{Shift towards Simulation-based Methods}
Simulation has become a central component of ADS debugging, reflecting its growing importance across both academia and industry. Although early simulation-based work was often motivated by the accessibility of lab environments with limited industrial funding, simulation is equally essential in industrial settings because of the high cost, safety risks, and logistical difficulties of debugging faults on real-world autonomous vehicles. As a result, simulation-based debugging is emerging as a practical and scalable approach for validating proof-of-concept ideas and conducting controlled, repeatable experiments. This trend mirrors broader developments in ADS testing, where simulation has become an increasingly dominant approach \cite{Tang2023}.

High demand for simulators in the field poses challenges due to their inherent limitations. \citet{testing-review} discusses some of the limitations that simulators face in testing, which are also shared by debugging approaches. Simulators remain advantageous because debugging approaches frequently leverage testing techniques which are easily applied and reproduced in simulation, allowing researchers to rely on well-established methodologies. Consequently, simulation-based studies provide a well-resourced foundation for exploring ADS debugging.

\subsubsection{Focus on Root Cause and Causality} \label{causality discussion}
Identifying the root cause of failure is a central goal of ADS debugging, as it enables developers to determine what must be changed to correct a system’s behaviour. Whereas general software systems can be repaired using manual debugging methods, the scale and complexity of ADS prohibit most manual techniques because they are too time-consuming; for example, even experienced developers require substantial time to follow a single execution trace \pref{P10}. 

Traditional debugging methods that rely on test coverage or execution traces are less effective for ADS because of its modular architecture. In ADS, most modules execute during every scenario, meaning a system trace typically covers a large portion of the source code and fails to localise a fault. Moreover, failed tests rarely halt execution in a way that reveals the exact failure point, such as a system crash. These limitations make it clear that conventional software debugging metrics do not apply to ADS. Therefore, recent research has focused on identifying causal relationships between ADS modules and their actions \pref{P10, P13} rather than on traditional metrics.

Simulation-based testing and debugging enable researchers to observe both the internal state of ADS modules and the external environment in a controlled, repeatable setting. This fine-grained visibility makes simulation a practical foundation for causality analysis, where the goal is to determine how specific module behaviours contribute to an observed system-level failure. Causal relationships between the actions of ADS and its modules can bridge the ``oracle gap'' \pref{P10}, meaning that existing ``many to one'' test failures at the scenario level can be turned into ``many to many'' or ``one to one'' at the module level.

Once a failing module is identified, traditional debugging techniques can again become effective. With the module-level output known, developers can either manually inspect the relevant code or apply established localisation methods. For example, the evaluation in \pref{P10} demonstrates how the Tarantula technique can be used to the implicated module to pinpoint faulty code lines.

\subsubsection{How Simplification Aids the Debugging Effort}
In this paper, we propose simplification as a distinct debugging task in the context of ADS. The complexity of ADS means that most techniques still require a simplification step to enable localisation tasks, as evidenced by the number of papers that include simplification techniques in our taxonomy (Figure \ref{fig:problem-taxonomy}). The increase in simulation-based validation studies enables the application of simplification techniques to the problem. Removing scenario elements is a common method of simplification \pref{P7, P11} that preserves fault triggers in the environment to facilitate root-cause identification and reduces processing load for repeat testing. A developer seeking to debug ADS should implement a method of scenario reduction, as the primary barrier to existing localisation techniques for standard systems is the complexity of ADS, which scenario reduction mitigates. Developers should also be aware that simplification is not a trivial task due to the ``ripping effect'' described in \pref{P7}. It states that removing active elements from the environment, such as cars, cannot be done as simply as removing those that do not directly interact with the ego vehicle, because interactions among scenario elements can affect how an element interacts with the ego. For example, one non-ego vehicle may delay another non-ego vehicle, causing it to interact with the ego at a later time. Existing techniques focus on active elements such as cars \pref{P7}, whereas future simplification techniques could reduce all aspects of the environment, including pedestrians and road elements. Identifying the specific fault triggers in the environment enables developers to remove them to achieve a passing scenario. Whilst this does not repair ADS, it does allow differential analysis between the passing and failing scenarios by observing ADS system states in both \pref{P11}. More debugging implementations could utilise fault triggers to identify internal code faults.

\subsubsection{Relationship between DNN Debugging and ADS Debugging}\label{sec:discussion:ml-debugging}
How DNNs are integrated and debugged, as well as the trend in XAI for ADS DNNs.

Deep learning is required for ADS with DNN models implemented within multiple ADS modules for prediction tasks. DNNs are regularly employed at the perception stage of ADS, with visual classifiers used to identify aspects of the environment \pref{P8}, and at the planning stage, using perception data to predict trajectories in the environment \pref{P10}. Where DNN debugging for ADS differs from standard machine learning is that the DNN is part of a much larger system, which will generally be debugged as a whole. Implementation of the previously discussed ADS debugging methods is required to localise the fault to the planning or prediction modules of the ADS, and, further, to the classifier output. When an ADS debugging method reaches a classifier, the incorrect output must be utilised with a generic DNN debugging method. Isolating the state of the DNN at failure is key to repair, for which existing tools \pref{P10, P13} are available. 

Once the conditions for debugging a DNN are met, the problem shifts to a generic classifier debugging issue, although there are still some influences from the nature of ADS. DNNs for ADS must operate in safety-critical settings, where misclassification can lead to safety violations. Repair methods need to be specialised for safety-critical DNNs, with greater emphasis on fine-tuning classifiers to yield precise results rather than merely retraining. Methods such as those proposed by \citet{paper13} are applicable in an ADS setting, as they focus on repairing specific classification failures without degrading other tasks. Risk levels for any particular misclassification are set in advance, so the repair tool can ensure that the most safety-critical issues are addressed first. Changing classifier weights can also affect classifications of other inputs, meaning that repairing one problem may shift the failure to another scenario. An adaptive repair method may be required to ensure that safety-critical failures are addressed. AdRep \cite{paper11} is a tool designed to repair weights of a DNN whilst actively re-evaluating the suspiciousness of each weight, so the issue of fault shifting is addressed. Both methods presented ensure that specific repairs are applied to ADS to ensure their safety.

If a developer seeks to debug DNNs within ADS, they should be aware of the two main points discussed regarding how the task differs from standard DNN debugging. Before debugging the specific input, output and state of the DNN, it needs to be found to be the source of the issue and recorded. This can then be followed by readily available repair techniques designed specifically for safety-critical DNNs, as standard retraining methods may introduce vulnerabilities into the system.

\subsection{Future Research Directions}

\subsubsection{Non-Unique Causality}
One of the biggest challenges in ADS debugging is that different root causes can yield similar, if not identical, outcomes. For example, if an ego vehicle fails to stop and collides with another car, the fault could be the perception module failing to detect the vehicle ahead, or a misconfigured control module failing to send the signal to activate the brakes. Debugging tools can struggle to distinguish the causes of faults given such limited system observability and output. Error propagation across modules complicates resolution, as incorrect output from the control module may result from less severe upstream failures. Current state-of-the-art causality analysis tools \pref{P10, P13} focus on identifying the single most likely cause of failure at a given moment in a scenario. As discussed in the work of \pref{P10}, the assumption that a failure emerges from a single module poses a current limitation despite the probability of multiple failure causes being statistically small. Multiple independent violation-inducing ADS modules could be further explored and may aid the general debugging of ADS systems, as simultaneous module analysis could increase the efficiency of existing causality analysis. Future work in ADS debugging could also utilise automatic causality analysis for automated program repair, as proposed by \pref{P13}. As previously discussed, automated repair methods require the automatic identification of faults and the generation of new code. Producing an automated repair tool from causality analysis would first require the development of a code-level localisation tool that utilises the results of causality analysis as a guide. New code to replace the identified code will then need to be generated, likely via mutation or a genetic algorithm.

\subsubsection{Further development in Root Cause Analysis and Repair}

In ADS development, we still lack an automated tool that can pinpoint the lines of code responsible for a given ADS failure. Current tools focus on module-level fault attribution \pref{P10} or identifying the specific misconfiguration in the system \pref{P11}. Whilst these existing tools are useful for assisting developers by reducing manual debugging effort, human intervention is still required. In addition to time spent on manual debugging, automated repair techniques cannot be applied without first achieving automated localisation. Automated code line localisation is the most straightforward approach to ADS debugging and could leverage conventional code localisation techniques, such as Tarantula \cite{tarantula}. As discussed, the scale of ADS blocks most conventional techniques from being applied, although current ADS localisation techniques could be combined to reduce the search space sufficiently for conventional techniques to be viable. Finding specific code faults within the ADS would remain challenging due to the number of code lines affected. In \pref{P4}, the average number of lines involved in ADS code repair was 104, indicating that conventional techniques may require modification to address multiple lines of code when there are many points of failure. Achieving automated localisation is the most challenging step currently encountered in ADS debugging. However, its implementation would simplify the debugging process and enable the application of existing line-repair techniques to ADS, thereby fully automating the debugging cycle.

\subsubsection{Debugging and Mitigation for Critical Modules}
ADS comprises many system modules, each of which is safety-critical. Debugging ADS is difficult, as accounting for all potential safety violations in such complex systems requires an as-yet unattained level of automatic fault localisation. There is an alternative approach that may be more achievable in the short term: fault mitigation within ADS. The work of \pref{P2} discusses how mitigation techniques can be applied to ADS to prevent safety-critical failures. One example discussed is the potential introduction of a safety layer alongside the planning module within ADS. This safety layer would contain a secondary trajectory planner, independent of the main one within the planning module, used for comparison before sending to control. Parameters for assessing trajectory feasibility can be set per ADS; if a trajectory is infeasible, the last calculated safe trajectory can be passed to control instead. The development of an additional "supervisor" module for ADS is also proposed, which would monitor the independent outputs of system modules and compare them against established safety standards. If a module deviates from the safe range, the supervisor can flag the system state as unsafe and trigger emergency action. Whilst fault mitigation isn't explicitly a debugging task, it is beneficial for future developers working on ADS to consider additional safety layers to prevent failures and address the ADS fault-propagation problem.

\subsubsection{Propose General Frameworks for ADS Debugging}

Earlier in this study (Section \ref{sec:problems}), we proposed definitions of the four debugging problems, tailored explicitly to ADS, to provide general terminology for future debugging work. The lack of general debugging terms such as these makes comparisons across papers difficult and makes the specifics of a study hard to understand. More general frameworks that provide clear methodologies would help focus the field, such as \pref{P2}, which offers a reproducible method for crash analysis of ADS. Future studies could focus on ADS failures beyond crashes, such as traffic-law violations or other forms of unsafe operation, to provide a helpful starting point for studies addressing more complex issues. Future ADS studies should also focus on the specific debugging task (s) they address, making it easier to understand what they aim to achieve. A future study could explicitly specify whether it is reaching, simplifying, localising, explaining, or repairing in relation to a problem, and which parts of its approach address each. An example from the literature is \pref{P11}, which explicitly mentions that simplification is used in the first step within the environment, followed by localisation within the ADS to locate the misconfiguration.

%% file: Sections/Conclusion.tex
\section{Conclusion}\label{sec:conclusion}

This systematic mapping study provides a structured overview of the current state of ADS debugging. Different approaches to problems were discussed, showing the range of methods, tools and techniques applied in the field. Specific definitions for each problem category were also proposed as a part of our effort to standardise terminology and standards for future work. We also provided a discussion on notable trends in the field, discussing challenges and opportunities for future work.

Collected results revealed that the field of ADS debugging is still in the early stages, with a small, but increasing, number of publications in recent years. Our analysis shows how many studies are in the early explorative stages, and we receive slow input from industry as ADS moves towards general use. We put specific emphasis on opportunities for future work and take note of promising directions current papers are moving in, such as bridging the gap from scenario triggers to internal faults. We hope this work inspires more contributions to debugging efforts in ADS and provides a strong starting point.

%% file: references.bib
@dataset{SMSdataset,
author = {Nathan Shaw and Sanjeetha Pennada and Robert M. Hierons and Donghwan Shin},
title = {ADS Debugging SMS Data Extraction Results},
year = {2025},
publisher = {ORDA},
doi = {10.15131/shef.data.29365220},
url = {https://doi.org/10.15131/shef.data.29365220},
organization = {University of Sheffield}
}

@Article{paper1,
AUTHOR = {Betz, Johannes and Heilmeier, Alexander and Wischnewski, Alexander and Stahl, Tim and Lienkamp, Markus},
TITLE = {Autonomous Driving—A Crash Explained in Detail},
JOURNAL = {Applied Sciences},
VOLUME = {9},
YEAR = {2019},
NUMBER = {23},
ARTICLE-NUMBER = {5126},
URL = {https://www.mdpi.com/2076-3417/9/23/5126},
ISSN = {2076-3417},
DOI = {10.3390/app9235126},
numpages = {23}
}

@ARTICLE{paper2,
  author={Fang, Yukun and Min, Haigen and Wang, Wuqi and Xu, Zhigang and Zhao, Xiangmo},
  journal={IEEE Sensors Journal}, 
  title={A Fault Detection and Diagnosis System for Autonomous Vehicles Based on Hybrid Approaches}, 
  year={2020},
  volume={20},
  number={16},
  pages={9359-9371},
  doi={10.1109/JSEN.2020.2987841}}

@INPROCEEDINGS{paper3,
  author={Orf, Stefan and Ochs, Sven and Doll, Jens and Schotschneider, Albert and Heinrich, Marc and Zofka, Marc René and Zöllner, J. Marius},
  booktitle={2024 IEEE 20th International Conference on Intelligent Computer Communication and Processing (ICCP)}, 
  title={Modular Fault Diagnosis Framework for Complex Autonomous Driving Systems}, 
  year={2024},
  volume={},
  number={},
  pages={1-8},
  doi={10.1109/ICCP63557.2024.10793003},
  address={Cluj-Napoca, Romania},
  publisher={IEEE}
}

@inproceedings{paper4,
author = {Abdessalem, Raja Ben and Panichella, Annibale and Nejati, Shiva and Briand, Lionel C. and Stifter, Thomas},
title = {Automated repair of feature interaction failures in automated driving systems},
year = {2020},
isbn = {9781450380089},
publisher = {Association for Computing Machinery},
address = {New York, NY, USA},
url = {https://doi.org/10.1145/3395363.3397386},
doi = {10.1145/3395363.3397386},
booktitle = {Proceedings of the 29th ACM SIGSOFT International Symposium on Software Testing and Analysis},
pages = {88–100},
numpages = {13},
location = {Virtual Event, USA},
series = {ISSTA 2020}
}

@inproceedings{paper5,
author = {Garcia, Joshua and Feng, Yang and Shen, Junjie and Almanee, Sumaya and Xia, Yuan and Chen, Qi Alfred},
title = {A comprehensive study of autonomous vehicle bugs},
year = {2020},
isbn = {9781450371216},
publisher = {Association for Computing Machinery},
address = {New York, NY, USA},
url = {https://doi.org/10.1145/3377811.3380397},
doi = {10.1145/3377811.3380397},
booktitle = {Proceedings of the ACM/IEEE 42nd International Conference on Software Engineering},
pages = {385–396},
numpages = {12},
location = {Seoul, South Korea},
series = {ICSE '20}
}

@INPROCEEDINGS{paper6,
  author={Arcaini, Paolo and Calò, Alessandro and Ishikawa, Fuyuki and Laurent, Thomas and Zhang, Xiao-Yi and Ali, Shaukat and Hauer, Florian and Ventresque, Anthony},
  booktitle={2021 IEEE Intelligent Vehicles Symposium Workshops (IV Workshops)}, 
  title={Parameter-Based Testing and Debugging of Autonomous Driving Systems}, 
  year={2021},
  volume={},
  number={},
  pages={197-202},
  doi={10.1109/IVWorkshops54471.2021.9669254},
  publisher={IEEE},
  address={Nagoya, Japan}
}

@article{paper7,
title = {Debugging Autonomous Driving Systems Using Serialized Software Components},
journal = {IFAC-PapersOnLine},
volume = {49},
number = {15},
pages = {44-49},
year = {2016},
note = {9th IFAC Symposium on Intelligent Autonomous Vehicles IAV 2016},
issn = {2405-8963},
doi = {https://doi.org/10.1016/j.ifacol.2016.07.612},
url = {https://www.sciencedirect.com/science/article/pii/S2405896316308837},
author = {Pascal Minnerup and David Lenz and Tobias Kessler and Alois Knoll}
}

@ARTICLE{paper8,
  author={Li, Meng and Wang, Yulei and Sun, Hengyang and Cui, Zhihao and Huang, Yanjun and Chen, Hong},
  journal={IEEE Transactions on Intelligent Vehicles}, 
  title={Explaining a Machine-Learning Lane Change Model With Maximum Entropy Shapley Values}, 
  year={2023},
  volume={8},
  number={6},
  pages={3620-3628},
  doi={10.1109/TIV.2023.3266196}}

@inproceedings{paper9,
author = {Feng, Shiwei and Ye, Yapeng and Shi, Qingkai and Cheng, Zhiyuan and Xu, Xiangzhe and Cheng, Siyuan and Choi, Hongjun and Zhang, Xiangyu},
title = {ROCAS: Root Cause Analysis of Autonomous Driving Accidents via Cyber-Physical Co-mutation},
year = {2024},
isbn = {9798400712487},
publisher = {Association for Computing Machinery},
address = {New York, NY, USA},
url = {https://doi.org/10.1145/3691620.3695530},
doi = {10.1145/3691620.3695530},
booktitle = {Proceedings of the 39th IEEE/ACM International Conference on Automated Software Engineering},
pages = {1620–1632},
numpages = {13},
location = {Sacramento, CA, USA},
series = {ASE '24}
}

@inproceedings{paper11,
author = {Li Calsi, Davide and Duran, Matias and Laurent, Thomas and Zhang, Xiao-Yi and Arcaini, Paolo and Ishikawa, Fuyuki},
title = {Adaptive Search-based Repair of Deep Neural Networks},
year = {2023},
isbn = {9798400701191},
publisher = {Association for Computing Machinery},
address = {New York, NY, USA},
url = {https://doi.org/10.1145/3583131.3590477},
doi = {10.1145/3583131.3590477},
booktitle = {Proceedings of the Genetic and Evolutionary Computation Conference},
pages = {1527–1536},
numpages = {10},
location = {Lisbon, Portugal},
series = {GECCO '23}
}

@INPROCEEDINGS{paper13,
  author={Calsi, Davide Li and Duran, Matias and Zhang, Xiao-Yi and Arcaini, Paolo and Ishikawa, Fuyuki},
  booktitle={2023 IEEE Conference on Software Testing, Verification and Validation (ICST)}, 
  title={Distributed Repair of Deep Neural Networks}, 
  year={2023},
  volume={},
  number={},
  pages={83-94},
  doi={10.1109/ICST57152.2023.00017},
  publisher={IEEE},
  address={Dublin, Ireland}
}

@INPROCEEDINGS{paper14,
  author={Roberts, Andrew and Heidari Iman, Mohammad Reza and Bellone, Mauro and Ghasempouri, Tara and Raik, Jaan and Maennel, Olaf and Hamad, Mohammad and Steinhorst, Sebastian},
  booktitle={2024 Design, Automation \& Test in Europe Conference \& Exhibition (DATE)}, 
  title={ADAssure: Debugging Methodology for Autonomous Driving Control Algorithms}, 
  year={2024},
  volume={},
  number={},
  pages={1-6},
  doi={10.23919/DATE58400.2024.10546519},
  publisher={IEEE},
  address={Valencia, Spain }
}

@inproceedings{paper15,
author = {Sun, Huijia and Poskitt, Christopher M. and Sun, Yang and Sun, Jun and Chen, Yuqi},
title = {ACAV: A Framework for Automatic Causality Analysis in Autonomous Vehicle Accident Recordings},
year = {2024},
isbn = {9798400702174},
publisher = {Association for Computing Machinery},
address = {New York, NY, USA},
url = {https://doi.org/10.1145/3597503.3639175},
doi = {10.1145/3597503.3639175},
booktitle = {Proceedings of the IEEE/ACM 46th International Conference on Software Engineering},
articleno = {102},
numpages = {13},
location = {Lisbon, Portugal},
series = {ICSE '24}
}

@INPROCEEDINGS{paper16,
  author={Arcaini, Paolo and Zhang, Xiao-Yi and Ishikawa, Fuyuki},
  booktitle={2022 IEEE Conference on Software Testing, Verification and Validation (ICST)}, 
  title={Less is More: Simplification of Test Scenarios for Autonomous Driving System Testing}, 
  year={2022},
  volume={},
  number={},
  pages={279-290},
  doi={10.1109/ICST53961.2022.00037},
  publisher={IEEE},
  address={Valencia, Spain}
}

@Article{paper18,
AUTHOR = {Hou, Wenkui and Li, Wanyu and Li, Pengyu},
TITLE = {Fault Diagnosis of the Autonomous Driving Perception System Based on Information Fusion},
JOURNAL = {Sensors},
VOLUME = {23},
YEAR = {2023},
NUMBER = {11},
ARTICLE-NUMBER = {5110},
URL = {https://www.mdpi.com/1424-8220/23/11/5110},
PubMedID = {37299837},
ISSN = {1424-8220},
DOI = {10.3390/s23115110},
PAGES = {5110},
}

@misc{paper20,
      title={Towards Automated Driving Violation Cause Analysis in Scenario-Based Testing for Autonomous Driving Systems}, 
      author={Ziwen Wan and Yuqi Huai and Yuntianyi Chen and Joshua Garcia and Qi Alfred Chen},
      year={2024},
      eprint={2401.10443},
      archivePrefix={arXiv},
      primaryClass={cs.SE},
      url={https://arxiv.org/abs/2401.10443}, 
}

@article{paper22,
author = {Chen, Yuntianyi and Huai, Yuqi and He, Yirui and Li, Shilong and Hong, Changnam and Chen, Qi Alfred and Garcia, Joshua},
title = {A Comprehensive Study of Bug-Fix Patterns in Autonomous Driving Systems},
year = {2025},
issue_date = {July 2025},
publisher = {Association for Computing Machinery},
address = {New York, NY, USA},
volume = {2},
number = {FSE},
url = {https://doi.org/10.1145/3715733},
doi = {10.1145/3715733},
journal = {Proc. ACM Softw. Eng.},
month = jun,
articleno = {FSE018},
numpages = {23}
}

@misc{zhong2021,
      title={A Survey on Scenario-Based Testing for Automated Driving Systems in High-Fidelity Simulation}, 
      author={Ziyuan Zhong and Yun Tang and Yuan Zhou and Vania de Oliveira Neves and Yang Liu and Baishakhi Ray},
      year={2021},
      eprint={2112.00964},
      archivePrefix={arXiv},
      primaryClass={cs.SE},
      url={https://arxiv.org/abs/2112.00964}, 
}

@ARTICLE{Sun2022,
  author={Sun, Jian and Zhang, He and Zhou, Huajun and Yu, Rongjie and Tian, Ye},
  journal={IEEE Transactions on Intelligent Transportation Systems}, 
  title={Scenario-Based Test Automation for Highly Automated Vehicles: A Review and Paving the Way for Systematic Safety Assurance}, 
  year={2022},
  volume={23},
  number={9},
  pages={14088-14103},
  doi={10.1109/TITS.2021.3136353}}

@INPROCEEDINGS{Schütt2023,
  author={Schütt, Barbara and Ransiek, Joshua and Braun, Thilo and Sax, Eric},
  booktitle={2023 IEEE Intelligent Vehicles Symposium (IV)}, 
  title={1001 Ways of Scenario Generation for Testing of Self-driving Cars: A Survey}, 
  year={2023},
  volume={},
  number={},
  pages={1-8},
  doi={10.1109/IV55152.2023.10186735},
  publisher={IEEE},
  address={Anchorage, Alaska, USA}
}

@article{Zahra2022,
author = {Sadri-Moshkenani, Zahra and Bradley, Justin and Rothermel, Gregg},
title = {Survey on test case generation, selection and prioritization for cyber-physical systems},
journal = {Software Testing, Verification and Reliability},
volume = {32},
number = {1},
pages = {e1794},
doi = {https://doi.org/10.1002/stvr.1794},
url = {https://onlinelibrary.wiley.com/doi/abs/10.1002/stvr.1794},
eprint = {https://onlinelibrary.wiley.com/doi/pdf/10.1002/stvr.1794},
year = {2022}
}

@inproceedings{Lou2022,
author = {Lou, Guannan and Deng, Yao and Zheng, Xi and Zhang, Mengshi and Zhang, Tianyi},
title = {Testing of autonomous driving systems: where are we and where should we go?},
year = {2022},
isbn = {9781450394130},
publisher = {Association for Computing Machinery},
address = {New York, NY, USA},
url = {https://doi.org/10.1145/3540250.3549111},
doi = {10.1145/3540250.3549111},
booktitle = {Proceedings of the 30th ACM Joint European Software Engineering Conference and Symposium on the Foundations of Software Engineering},
pages = {31–43},
numpages = {13},
location = {Singapore, Singapore},
series = {ESEC/FSE 2022}
}

@ARTICLE{Ding2023,
  author={Ding, Wenhao and Xu, Chejian and Arief, Mansur and Lin, Haohong and Li, Bo and Zhao, Ding},
  journal={IEEE Transactions on Intelligent Transportation Systems}, 
  title={A Survey on Safety-Critical Driving Scenario Generation—A Methodological Perspective}, 
  year={2023},
  volume={24},
  number={7},
  pages={6971-6988},
  doi={10.1109/TITS.2023.3259322}}

@article{Corso_2021,
   title={A Survey of Algorithms for Black-Box Safety Validation of Cyber-Physical Systems},
   volume={72},
   ISSN={1076-9757},
   url={http://dx.doi.org/10.1613/jair.1.12716},
   DOI={10.1613/jair.1.12716},
   journal={Journal of Artificial Intelligence Research},
   publisher={AI Access Foundation},
   author={Corso, Anthony and Moss, Robert and Koren, Mark and Lee, Ritchie and Kochenderfer, Mykel},
   year={2021},
   month={oct},
   numpages={51}
}

@misc{birchler2024,
      title={A Roadmap for Simulation-Based Testing of Autonomous Cyber-Physical Systems: Challenges and Future Direction}, 
      author={Christian Birchler and Sajad Khatiri and Pooja Rani and Timo Kehrer and Sebastiano Panichella},
      year={2024},
      eprint={2405.01064},
      archivePrefix={arXiv},
      primaryClass={cs.SE},
      url={https://arxiv.org/abs/2405.01064}, 
}

@article{Araujo,
author = {Araujo, Hugo and Mousavi, Mohammad Reza and Varshosaz, Mahsa},
title = {Testing, Validation, and Verification of Robotic and Autonomous Systems: A Systematic Review},
year = {2023},
issue_date = {March 2023},
publisher = {Association for Computing Machinery},
address = {New York, NY, USA},
volume = {32},
number = {2},
issn = {1049-331X},
url = {https://doi.org/10.1145/3542945},
doi = {10.1145/3542945},
journal = {ACM Trans. Softw. Eng. Methodol.},
month = mar,
articleno = {51},
numpages = {61}
}

@article{Tang2023,
author = {Tang, Shuncheng and Zhang, Zhenya and Zhang, Yi and Zhou, Jixiang and Guo, Yan and Liu, Shuang and Guo, Shengjian and Li, Yan-Fu and Ma, Lei and Xue, Yinxing and Liu, Yang},
title = {A Survey on Automated Driving System Testing: Landscapes and Trends},
year = {2023},
issue_date = {September 2023},
publisher = {Association for Computing Machinery},
address = {New York, NY, USA},
volume = {32},
number = {5},
issn = {1049-331X},
url = {https://doi.org/10.1145/3579642},
doi = {10.1145/3579642},
journal = {ACM Trans. Softw. Eng. Methodol.},
month = jul,
articleno = {124},
numpages = {62}
}

@ARTICLE{Zhang2023,
  author={Zhang, Xinhai and Tao, Jianbo and Tan, Kaige and Törngren, Martin and Sánchez, José Manuel Gaspar and Ramli, Muhammad Rusyadi and Tao, Xin and Gyllenhammar, Magnus and Wotawa, Franz and Mohan, Naveen and Nica, Mihai and Felbinger, Hermann},
  journal={IEEE Transactions on Software Engineering}, 
  title={Finding Critical Scenarios for Automated Driving Systems: A Systematic Mapping Study}, 
  year={2023},
  volume={49},
  number={3},
  pages={991-1026},
  doi={10.1109/TSE.2022.3170122}}

@article{PETERSEN20151,
title = {Guidelines for conducting systematic mapping studies in software engineering: An update},
journal = {Information and Software Technology},
volume = {64},
pages = {1-18},
year = {2015},
issn = {0950-5849},
doi = {https://doi.org/10.1016/j.infsof.2015.03.007},
url = {https://www.sciencedirect.com/science/article/pii/S0950584915000646},
author = {Kai Petersen and Sairam Vakkalanka and Ludwik Kuzniarz}
}

@article{BRERETON2007571,
title = {Lessons from applying the systematic literature review process within the software engineering domain},
journal = {Journal of Systems and Software},
volume = {80},
number = {4},
pages = {571-583},
year = {2007},
note = {Software Performance},
issn = {0164-1212},
doi = {https://doi.org/10.1016/j.jss.2006.07.009},
url = {https://www.sciencedirect.com/science/article/pii/S016412120600197X},
author = {Pearl Brereton and Barbara A. Kitchenham and David Budgen and Mark Turner and Mohamed Khalil}
}

@misc{Priem2022OpenAlex,
  author    = {Priem, Jason and Piwowar, Heather and Orr, Richard},
  title     = {OpenAlex: A fully-open index of scholarly works, authors, venues, institutions, and concepts},
  journal   = {ArXiv},
  year      = {2022},
  url       = {https://arxiv.org/abs/2205.01833},
  archivePrefix = {arXiv},
  eprint    = {2205.01833}
}

@misc{Alperin2024AnAO,
      title={An analysis of the suitability of OpenAlex for bibliometric analyses}, 
      author={Juan Pablo Alperin and Jason Portenoy and Kyle Demes and Vincent Larivière and Stefanie Haustein},
      year={2024},
      eprint={2404.17663},
      archivePrefix={arXiv},
      primaryClass={cs.DL},
      url={https://arxiv.org/abs/2404.17663}, 
}

@ARTICLE{Yurtsever,
  author={Yurtsever, Ekim and Lambert, Jacob and Carballo, Alexander and Takeda, Kazuya},
  journal={IEEE Access}, 
  title={A Survey of Autonomous Driving: Common Practices and Emerging Technologies}, 
  year={2020},
  volume={8},
  number={},
  pages={58443-58469},
  doi={10.1109/ACCESS.2020.2983149}}

@misc{gov,
	title = {Self-driving vehicles set to be on roads by 2026 as {Automated} {Vehicles} {Act} becomes law},
	url = {https://www.gov.uk/government/news/self-driving-vehicles-set-to-be-on-roads-by-2026-as-automated-vehicles-act-becomes-law},
	language = {en},
	urldate = {2025-05-17},
	journal = {GOV.UK},
        year         = {2024},
        month        = {May},
        day          = {20},
        author       = {{Department for Transport} and {Centre for Connected and Autonomous Vehicles} and {The Rt Hon Mark Harper}},
}

@article{Koopman2016,
author = {Koopman, Philip and Wagner, Michael},
year = {2016},
month = {04},
pages = {15-24},
title = {Challenges in Autonomous Vehicle Testing and Validation},
volume = {4},
journal = {SAE International Journal of Transportation Safety},
doi = {10.4271/2016-01-0128}
}

@article{castelvecchi_can_2016,
	title = {Can we open the black box of {AI}?},
	volume = {538},
	url = {http://www.nature.com/news/can-we-open-the-black-box-of-ai-1.20731},
	doi = {10.1038/538020a},
	pages = {20},
	number = {7623},
	journaltitle = {Nature News},
        journal = {Nature},
	author = {Castelvecchi, Davide},
	urldate = {2025-05-19},
	date = {2016-10-06},
	langid = {english},
	note = {Cg\_type: Nature News
Section: News Feature},
    year = {2016}
}

@ARTICLE{fault-local,
  author={Wong, W. Eric and Gao, Ruizhi and Li, Yihao and Abreu, Rui and Wotawa, Franz},
  journal={IEEE Transactions on Software Engineering}, 
  title={A Survey on Software Fault Localization}, 
  year={2016},
  volume={42},
  number={8},
  pages={707-740},
  doi={10.1109/TSE.2016.2521368}}

@article{repair-survey,
author = {Monperrus, Martin},
title = {Automatic Software Repair: A Bibliography},
year = {2018},
issue_date = {January 2019},
publisher = {Association for Computing Machinery},
address = {New York, NY, USA},
volume = {51},
number = {1},
issn = {0360-0300},
url = {https://doi-org.sheffield.idm.oclc.org/10.1145/3105906},
doi = {10.1145/3105906},
journal = {ACM Comput. Surv.},
month = jan,
articleno = {17},
numpages = {24}
}

@ARTICLE{5386906,
  author={Hailpern, B. and Santhanam, P.},
  journal={IBM Systems Journal}, 
  title={Software debugging, testing, and verification}, 
  year={2002},
  volume={41},
  number={1},
  pages={4-12},
  doi={10.1147/sj.411.0004}}

@inproceedings{Wohlin2014,
author = {Wohlin, Claes},
title = {Guidelines for snowballing in systematic literature studies and a replication in software engineering},
year = {2014},
isbn = {9781450324762},
publisher = {Association for Computing Machinery},
address = {New York, NY, USA},
url = {https://doi-org.sheffield.idm.oclc.org/10.1145/2601248.2601268},
doi = {10.1145/2601248.2601268},
booktitle = {Proceedings of the 18th International Conference on Evaluation and Assessment in Software Engineering},
articleno = {38},
numpages = {10},
location = {London, England, United Kingdom},
series = {EASE '14}
}

@INPROCEEDINGS{CAV-paper,
  author={Khalil, Abdelrahman and Al Janaideh, Mohammad and Aljanaideh, Khaled F. and Kundur, Deepa},
  booktitle={2020 American Control Conference (ACC)}, 
  title={Fault Detection, Localization, and Mitigation of a Network of Connected Autonomous Vehicles Using Transmissibility Identification}, 
  year={2020},
  volume={},
  number={},
  pages={386-391},
  doi={10.23919/ACC45564.2020.9147801},
  publisher={IEEE},
  address={Denver, CO, USA}
}

@inproceedings{XAI-paper,
      title={Explaining Autonomous Driving by Learning End-to-End Visual Attention}, 
      author={Luca Cultrera and Lorenzo Seidenari and Federico Becattini and Pietro Pala and Alberto Del Bimbo},
      booktitle = {Proceedings of the CVPR Workshop on Safe Artificial Intelligence for Automated Driving (SAIAD)},
      year={2020},
      eprint={2006.03347},
      archivePrefix={arXiv},
      primaryClass={cs.CV},
      url={https://arxiv.org/abs/2006.03347}, 
      address={Seattle, Washington},
      publisher={IEEE / Computer Vision Foundation (CVF)},
      numpages={10}
}

@ARTICLE{delta-debug,
  author={Zeller, A. and Hildebrandt, R.},
  journal={IEEE Transactions on Software Engineering}, 
  title={Simplifying and isolating failure-inducing input}, 
  year={2002},
  volume={28},
  number={2},
  pages={183-200},
  doi={10.1109/32.988498}}

@inproceedings{Holger2005,
author = {Cleve, Holger and Zeller, Andreas},
title = {Locating causes of program failures},
year = {2005},
isbn = {1581139632},
publisher = {Association for Computing Machinery},
address = {New York, NY, USA},
url = {https://doi.org/10.1145/1062455.1062522},
doi = {10.1145/1062455.1062522},
booktitle = {Proceedings of the 27th International Conference on Software Engineering},
pages = {342–351},
numpages = {10},
location = {St. Louis, MO, USA},
series = {ICSE '05}
}

@INPROCEEDINGS{testing-review,
  author={Huang, WuLing and Kunfeng Wang and Yisheng Lv and FengHua Zhu},
  booktitle={2016 IEEE 19th International Conference on Intelligent Transportation Systems (ITSC)}, 
  title={Autonomous vehicles testing methods review}, 
  year={2016},
  volume={},
  number={},
  pages={163-168},
  doi={10.1109/ITSC.2016.7795548},
  publisher={IEEE},
  address={Rio de Janeiro, Brazil}
}

@book{Zeller2009,
author = {Zeller, Andreas},
title = {Why Programs Fail, Second Edition: A Guide to Systematic Debugging},
year = {2009},
isbn = {0123745152},
publisher = {Morgan Kaufmann Publishers Inc.},
address = {San Francisco, CA, USA},
edition = {2nd}
}

@inproceedings{SHAP2017,
author = {Lundberg, Scott M. and Lee, Su-In},
title = {A unified approach to interpreting model predictions},
year = {2017},
isbn = {9781510860964},
publisher = {Curran Associates Inc.},
address = {Red Hook, NY, USA},
booktitle = {Proceedings of the 31st International Conference on Neural Information Processing Systems},
pages = {4768–4777},
numpages = {10},
location = {Long Beach, California, USA},
series = {NIPS'17}
}

@article{Arachne,
author = {Sohn, Jeongju and Kang, Sungmin and Yoo, Shin},
title = {Arachne: Search-Based Repair of Deep Neural Networks},
year = {2023},
issue_date = {July 2023},
publisher = {Association for Computing Machinery},
address = {New York, NY, USA},
volume = {32},
number = {4},
issn = {1049-331X},
url = {https://doi.org/10.1145/3563210},
doi = {10.1145/3563210},
journal = {ACM Trans. Softw. Eng. Methodol.},
month = may,
articleno = {85},
numpages = {26}
}

@article{genetic-algorithm,
author = {Thede, Scott},
year = {2004},
month = {10},
pages = {},
title = {An introduction to genetic algorithms},
volume = {20},
journal = {Journal of Computing Sciences in Colleges}
}

@techreport{keele2007guidelines,
  title={Guidelines for performing systematic literature reviews in software engineering},
  author={Keele, Staffs and others},
  year={2007},
  institution={Technical report, ver. 2.3 ebse technical report. ebse}
}

@article{Algorithmic_debugging_survey,
author = {Caballero, Rafael and Riesco, Adri\'{a}n and Silva, Josep},
title = {A Survey of Algorithmic Debugging},
year = {2017},
issue_date = {July 2018},
publisher = {Association for Computing Machinery},
address = {New York, NY, USA},
volume = {50},
number = {4},
issn = {0360-0300},
url = {https://doi.org/10.1145/3106740},
doi = {10.1145/3106740},
journal = {ACM Comput. Surv.},
month = aug,
articleno = {60},
numpages = {35}
}

@ARTICLE{Automatic_software_repair_survey,
  author={Gazzola, Luca and Micucci, Daniela and Mariani, Leonardo},
  journal={IEEE Transactions on Software Engineering}, 
  title={Automatic Software Repair: A Survey}, 
  year={2019},
  volume={45},
  number={1},
  pages={34-67},
  doi={10.1109/TSE.2017.2755013}}

@inproceedings{Automated_debugging_helping,
author = {Parnin, Chris and Orso, Alessandro},
title = {Are automated debugging techniques actually helping programmers?},
year = {2011},
isbn = {9781450305624},
publisher = {Association for Computing Machinery},
address = {New York, NY, USA},
url = {https://doi.org/10.1145/2001420.2001445},
doi = {10.1145/2001420.2001445},
booktitle = {Proceedings of the 2011 International Symposium on Software Testing and Analysis},
pages = {199–209},
numpages = {11},
location = {Toronto, Ontario, Canada},
series = {ISSTA '11}
}

@ARTICLE{program_comprehension,
  author={Cornelissen, Bas and Zaidman, Andy and van Deursen, Arie and Moonen, Leon and Koschke, Rainer},
  journal={IEEE Transactions on Software Engineering}, 
  title={A Systematic Survey of Program Comprehension through Dynamic Analysis}, 
  year={2009},
  volume={35},
  number={5},
  pages={684-702},
  doi={10.1109/TSE.2009.28}}

@article{XAI-survey,
	title = {Explainability of {Deep} {Vision}-{Based} {Autonomous} {Driving} {Systems}: {Review} and {Challenges}},
	volume = {130},
	issn = {1573-1405},
	url = {https://doi.org/10.1007/s11263-022-01657-x},
	doi = {10.1007/s11263-022-01657-x},
	number = {10},
	journal = {International Journal of Computer Vision},
	author = {Zablocki, Éloi and Ben-Younes, Hédi and Pérez, Patrick and Cord, Matthieu},
	month = oct,
	year = {2022},
	pages = {2425--2452},
}

@inproceedings{tarantula,
author = {Jones, James A. and Harrold, Mary Jean and Stasko, John},
title = {Visualization of test information to assist fault localization},
year = {2002},
isbn = {158113472X},
publisher = {Association for Computing Machinery},
address = {New York, NY, USA},
url = {https://doi-org.sheffield.idm.oclc.org/10.1145/581339.581397},
doi = {10.1145/581339.581397},
booktitle = {Proceedings of the 24th International Conference on Software Engineering},
pages = {467–477},
numpages = {11},
location = {Orlando, Florida},
series = {ICSE '02}
}

@INPROCEEDINGS{11052713,
  author={Osikowicz, Olek and McMinn, Phil and Shin, Donghwan},
  booktitle={2025 IEEE/ACM International Flaky Tests Workshop (FTW)}, 
  title={Empirically Evaluating Flaky Tests for Autonomous Driving Systems in Simulated Environments}, 
  year={2025},
  volume={},
  number={},
  pages={13-20},
  doi={10.1109/FTW66604.2025.00009}}

@ARTICLE{9793395,
  author={Chance, Greg and Ghobrial, Abanoub and McAreavey, Kevin and Lemaignan, Séverin and Pipe, Tony and Eder, Kerstin},
  journal={IEEE Transactions on Intelligent Transportation Systems}, 
  title={On Determinism of Game Engines Used for Simulation-Based Autonomous Vehicle Verification}, 
  year={2022},
  volume={23},
  number={11},
  pages={20538-20552},
  doi={10.1109/TITS.2022.3177887}}
